\documentclass[preprint,12pt]{elsarticle}

\usepackage{lineno,hyperref}
\usepackage{bm}
\usepackage{amsfonts}
\usepackage{amssymb}
\usepackage{amsmath}
\usepackage{comment}
\usepackage{isomath}

\usepackage{adjustbox}

\usepackage{scalerel,stackengine}
\stackMath
\newcommand\reallywidehat[1]{%
\savestack{\tmpbox}{\stretchto{%
  \scaleto{%
    \scalerel*[\widthof{\ensuremath{#1}}]{\kern-.6pt\bigwedge\kern-.6pt}%
    {\rule[-\textheight/2]{1ex}{\textheight}}
  }{\textheight}%
}{0.5ex}}%
\stackon[1pt]{#1}{\tmpbox}%
}

\newcommand{\bbGamma}{{\mathpalette\makebbGamma\relax}}
\newcommand{\makebbGamma}[2]{%
  \raisebox{\depth}{\scalebox{1}[-1]{$\mathsurround=0pt#1\mathbb{L}$}}%
}

\usepackage{stmaryrd} 
\usepackage[usenames]{color}
\usepackage{epsfig}
\usepackage{graphicx}
\usepackage{psfrag}

\graphicspath{{Figures/}}
\usepackage{float}
\floatstyle{boxed}
\newfloat{algorithm}{htb}{loa}
\floatname{algorithm}{Algorithm}
\usepackage{algorithm,algpseudocode}
\usepackage{caption}
\usepackage{subcaption}
\usepackage{color}
\usepackage[table,xcdraw]{xcolor}
\usepackage{multirow}
\usepackage{natbib}
\modulolinenumbers[5]

\usepackage[left = 2cm, right=2.5cm, top=2cm, bottom =2cm]{geometry}

\begin{document}
\begin{frontmatter}

\title{SPiFOL: A Spectral-based Physics-informed Finite Operator Learning for Prediction of Mechanical Behavior of Microstructures}

\author{Ali Harandi$^{1*}$, Hooman Danesh$^1$, Kevin Linka$^{1}$, Stefanie Reese$^{1,\,2}$, Shahed Rezaei$^{3}$}
\address{$^1$Institute of Applied Mechanics, \\ RWTH Aachen University, Mies-van-der-Rohe-Str. 1, D-52074 Aachen, Germany}
\address{$^2$University of Siegen, Adolf-Reichwein-Str. 2a, D-57076 Siegen, Germany}
\address{$^3$ACCESS e.V., Intzestr. 5, D-52072 Aachen, Germany}
\address{$^*$ corresponding author: ali.harandi@rwth-aachen.de}

\begin{abstract}
A novel physics-informed operator learning technique based on spectral methods is introduced to model the complex behavior of heterogeneous materials. The Lippmann-Schwinger operator in Fourier space is employed to construct physical constraints with minimal computational overhead, effectively eliminating the need for automatic differentiation.
The introduced methodology accelerates the training process by enabling gradient construction on a fixed, finite discretization in Fourier space. Later, the spectral physics-informed finite operator learning (SPiFOL) framework is built based on this discretization and trained to map the arbitrary shape of microstructures to their mechanical responses (strain fields) without relying on labeled data. The training is done by minimizing equilibrium in Fourier space concerning the macroscopic loading condition, which also guarantees the periodicity.
SPiFOL, as a physics-informed operator learning method, enables rapid predictions through forward inference after training. To ensure accuracy, we incorporate physical constraints and diversify the training data. However, performance may still degrade for out-of-distribution microstructures.
SPiFOL is further enhanced by integrating a Fourier Neural Operator (FNO). Compared to the standard data-driven FNO, SPiFOL shows higher accuracy in predicting stress fields and provides nearly resolution-independent results. Additionally, its zero-shot super-resolution capabilities are explored in heterogeneous domains.
Finally, SPiFOL is extended to handle 3D problems and further adapted to finite elasticity, demonstrating the robustness of the framework in handling nonlinear mechanical behavior. The framework shows great potential for efficient and scalable prediction of mechanical responses in complex material systems while also reducing the training time required for training physics-informed neural operators.

\end{abstract} 

\end{frontmatter}
\section{Introduction} 

Modeling complex phenomena in engineering problems typically involves formulating and solving partial differential equations (PDEs). Numerical methods such as the Finite Element Method (FEM) \cite{liu2022eighty}, the Finite Difference Method (FDM) \cite{sod1978survey}, and spectral methods \cite{shen2011spectral} are commonly used to solve these PDEs. Despite their strong performance in handling complex systems of equations, these numerical methods must be applied repeatedly whenever problem parameters—such as initial and boundary conditions, source terms, or PDE coefficients—vary.

The same problem applies to certain deep learning (DL) techniques, such as physics-informed neural networks (PINNs). The core idea behind PINNs is to incorporate physical constraints, i.e., all conditions and requirements defined in a given boundary value problem, to evaluate solutions immediately after optimizing the neural network (NN) hyperparameters \cite{karniadakis2021physics}. For applications of PINNs in heterogeneous domains, see \cite{Rao2021, REZAEI2022PINN, Harandi2023}.

While PINNs offer significant advantages, such as real-time prediction and enabling inverse design \cite{Luinverse, Chen2020}, they often lack generalizability. Moreover, retraining them for new sets of PDE coefficients can be computationally expensive and may even fail to converge to a solution \cite{WANG2022why, krishnapriyan2021characterizing}.

Therefore, it is essential to approximate solution operators that map input functions to output functions (i.e., solutions to PDEs) using both data and physical principles. Building on the universal approximation theorem for operators, the Deep Operator Network (DeepONet) was introduced in \cite{Lu2021}.
DeepONet has been applied to a wide range of problems, including modeling plastic behavior under varying loading conditions and geometries \cite{koric2024deep, he2023novel}, as well as phase-field fracture problems with different initial notch positions \cite{GOSWAMI2022114587}. Several enhancements have been proposed to improve the accuracy of DeepONets \cite{wang2022improved, haghighat2024deeponet, LU2022114778, kontolati2024learning}. Additionally, \citet{he2024geom} and \citet{bahmani2024resolution} extended DeepONet to handle parametric geometries and resolution-independent input functions, respectively.

Another prominent class of neural operators is the Fourier Neural Operator (FNO), which performs kernel parameterization in Fourier space, as introduced in \cite{li2020neural, kovachki2023neural}. FNO leverages the Fast Fourier Transform (FFT), which restricts its application to rectangular domains. To overcome this limitation, \citet{li2023geofno} utilized a latent space representation to extend FNO’s capability to irregular domains.
FNO has also been employed in various applications. For example, \citet{Mehran2022} used FNO to map digital composite microstructures to their corresponding stress fields by incorporating labeled data, demonstrating FNO’s superiority over U-Net. \citet{YOU2022115296} generalized FNO for varying macroscopic loading conditions over time by introducing iterative loading layers into the FNO architecture, enabling the prediction of behavior under unseen loading scenarios. Furthermore, \citet{wang2024homogenius} applied FNO to develop data-driven surrogate models for triply periodic minimal surface (TPMS) metamaterials.

Comprehensive discussions on neural operator architectures are provided by \citet{kovachki2024operator}, while \citet{BOULLE202483} analyzed these models from the perspective of numerical algebra. Despite their promising capabilities, these approaches heavily rely on data, which often requires extensive offline numerical simulations to generate. While this data generation step is crucial for building accurate surrogate models, it is computationally expensive, limiting their scalability. Moreover, the performance of such models can be questionable, as they are typically supervised only by data within a limited range of scenarios.

Similar to the application of physical constraints in conventional neural networks (NNs) \cite{faroughi2022physics}, physical constraints can be incorporated either alongside data-based loss functions or even in their absence. In the context of DeepONet, \citet{wang2021learningphysdeepon} developed a physics-informed DeepONet for solving various types of partial differential equations (PDEs). Additionally, \citet{li2023phasedeeponet} proposed a phase-informed DeepONet to predict the dynamic response of systems by controlling the evolution of their free energy.

\citet{li2023physicsinformed} introduced a physics-informed FNO by embedding PDE-based loss functions within the model architecture. Furthermore, \citet{khorrami2024divergence} and \citet{kapoor2022comparison} conducted comparative studies on physics-informed and physics-encoded FNOs, evaluating their performance in combination with data-driven loss functions. \citet{RASHID2023105444} explored different neural operator architectures for predicting strain evolution in digital composites.

Comprehensive reviews of neural operator methodologies, with and without the incorporation of physical constraints, have been provided by \citet{GOSWAMI20TPF} and \citet{rosofsky2023applications}.

Despite significant advances in operator learning, embedding physics in a fast and robust manner remains a challenge. Computing the gradients required to build physical constraints via automatic differentiation (AD) is expensive, see \cite{CHIU2022114909}. To mitigate these shortcomings, several authors have attempted to combine conventional numerical methods to construct the gradients needed to build PDE constraints. \cite{REN2022114399, ZHAO2023105516, ZHANG2023111919} combined CNN approaches with FD and finite volume methods, see also \cite{Phillips2023}. 

To integrate FEM with operator learning, \citet{Rezaei2024fol_mech} used FEM to compute the gradients needed for the physical loss function of a neural network. This network maps microstructural features to mechanical deformations at fixed grid points discretized by FEM. Similarly, \citet{yamazaki2024} addressed the transient heat equation through an approach where a finite operator learning (FOL) framework maps the initial temperature field to the subsequent one by minimizing the residual form of the governing equation, which is also formulated using FEM.
Further advances were introduced by \citet{Rezaei2024finite}, who used a Sobolev training strategy to improve the proposed methodology for design applications. However, the trained networks in these approaches are constrained by predefined discretizations. To obtain solutions at arbitrary points within the domain, additional interpolation techniques are required.

The challenges associated with multiscale computational techniques in solid mechanics are one of the primary motivations for this work.
In micromechanics, homogenization aims to determine the effective behavior of heterogeneous microstructures across multiple scales. Methods such as FE² \cite{raju2021review} and FE-FFT \cite{SPAHN2014871} are among the most widely used approaches for two-scale computations. FFT-based techniques are particularly valued for their speed and accuracy in microscale simulations \cite{lucarini2021fft, schneider2021review}. Interestingly, the inherent limitations of FFT can be advantageous in multiscale simulations, where maintaining periodicity at the microscale is often a critical requirement.
\citet{Risthaus24} extended the FFT framework to accommodate Dirichlet boundary conditions, while \citet{lucarini2023fft} further advanced the method for modeling fatigue crack initiation. Furthermore, \citet{danesh2024fft} used FFT-based approaches to generate datasets for surrogate modeling of metamaterials. \citet{kumar2020assessment} demonstrated the superior performance of the FFT scheme over the Finite Element Method (FEM) in capturing microstructural behavior associated with nonconvex potentials.
\citet{schneider2019polarization} compared polarization-based schemes with gradient-based solvers for FFT-based computational homogenization of inelastic materials. While these methods are effective, their computational complexity remains significant, especially when dealing with microstructures with high phase contrast ratios. For each new microstructure and discretization, an FFT simulation must be performed, and the Lippmann-Schwinger operator - essential for enforcing physical constraints - must be recomputed, further increasing the computational burden. 
\begin{figure}[htb] 
  \centering
  \includegraphics[width=1.0\linewidth]{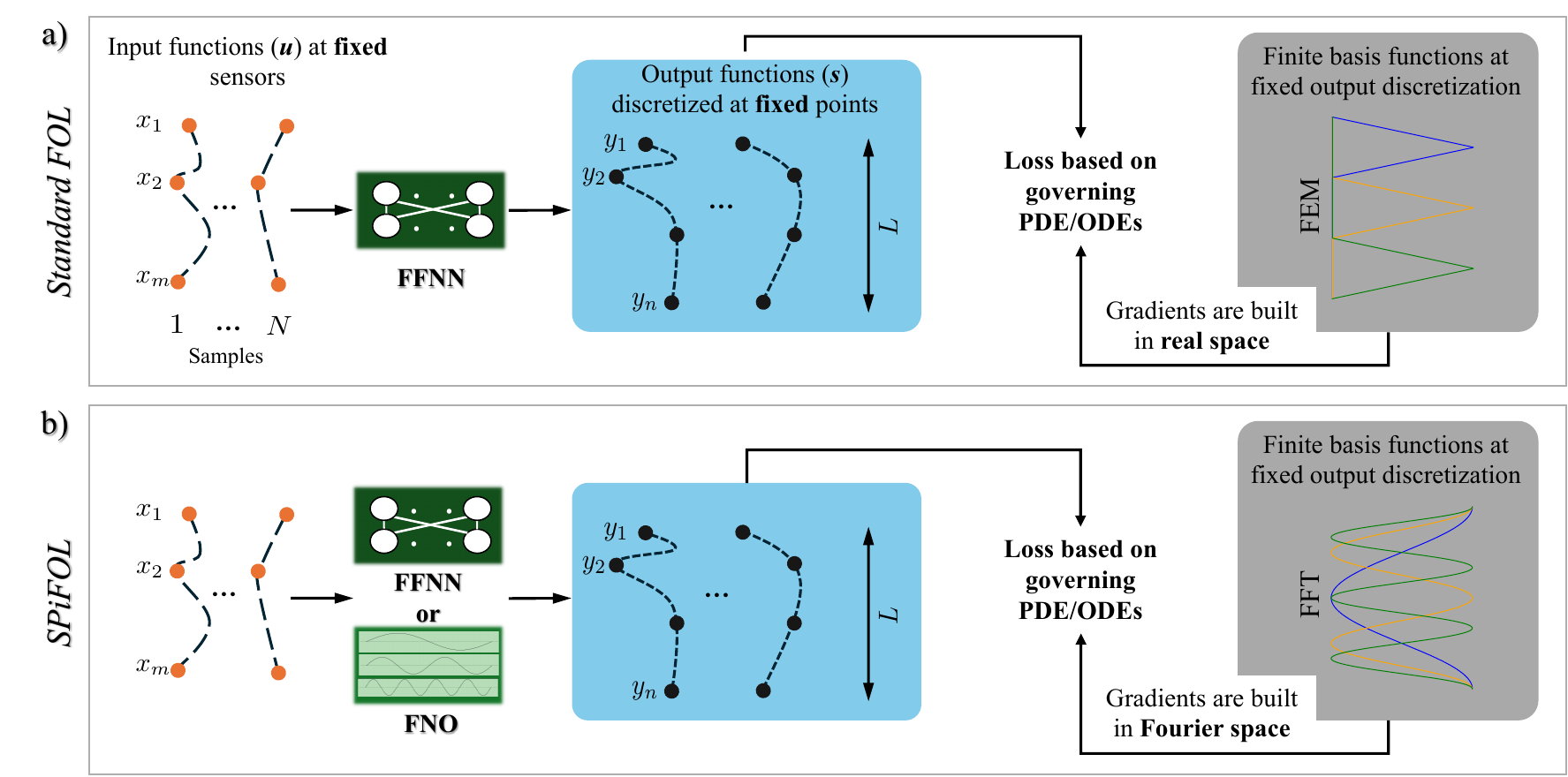}
  \caption{a) Standard FOL (\cite{Rezaei2024finite, Rezaei2024fol_mech}), where the input function is evaluated at sensors ($\bm{u}$) and mapped to the output function ($\bm{s}$) by feed-forward neural networks (FFNN). The network predicts the output function at the fixed points in the output space where gradients in real space are generated by numerical tools like FEM. b) SPiFOL maps the input function to the output function using FFNN or FNO neural architectures. The gradients of the output function are built in Fourier space, which is used to build physical constraints in the loss.}
\label{fig:intro}
\end{figure}

SPiFOL aims to develop physics-based operators that serve as surrogate models for parametric partial differential equations (PDEs), using the principles of FFT-based approaches in micromechanics. The physical equations are embedded directly in Fourier space, and the loss function is formulated with minimal additional computational effort by applying the Fourier transform and using the Lippmann-Schwinger operator, which is pre-computed in Fourier space prior to training. The loss function is derived similarly to conventional FFT-based methods, following the fixed-point scheme proposed by \cite{moulinec1998numerical}.
Fig.\,\ref{fig:intro}, shows how Fourier-based shape functions are employed at a fixed output finite space, similar to the standard FOL idea, to build gradients (in Fourier space) which are required to formulate physical loss. 
In addition, we use the FNO architecture, which provides the flexibility for SPiFOL to use different input resolutions and predict corresponding output solutions. The latter tackles a shortcoming of conventional FOL operators to predict the response for different input function resolutions. 
The novel developments of this work are summarized as follows:
\begin{itemize}
    \item Physics-informed training in Fourier space with almost no additional overhead compared to the conventional physics-informed neural networks in which PDEs are constructed by AD or other numerical methods which are computed in real space like FEM or FDM. 
    \item The network maps microstructure topology directly to its strain components in a purely unsupervised manner, two orders of magnitude faster than conventional FFT solvers. It is important to note that in many existing studies, only the primary variable, such as displacement (or temperature), is predicted. As a result, additional derivations are required to obtain strain values, which may introduce more errors in the predictions.
    The SPiFOL methodology is adapted to handle 3D microstructures and further extended for applications in finite elasticity. 
\end{itemize}

The structure of the paper is as follows: Section \ref{sec:spifol} provides a detailed overview of the SPiFOL methodology, explaining how SPiFOL maps different microstructures to their corresponding mechanical responses in the small deformation regime. For finite deformation, it describes the mapping of macroscopic deformation gradients applied to a microstructure to its resulting mechanical behavior. Finally, section \ref{sec:results} demonstrates the ability of the network to predict the mechanical behavior of a wide range of microstructures, followed by a conclusion and outlook on the future directions of this work.
\section{SPiFOL methodology for mapping microstructures to stresses}
\label{sec:spifol}
\begin{figure}[htb] 
  \centering
  \includegraphics[width=0.97\linewidth]{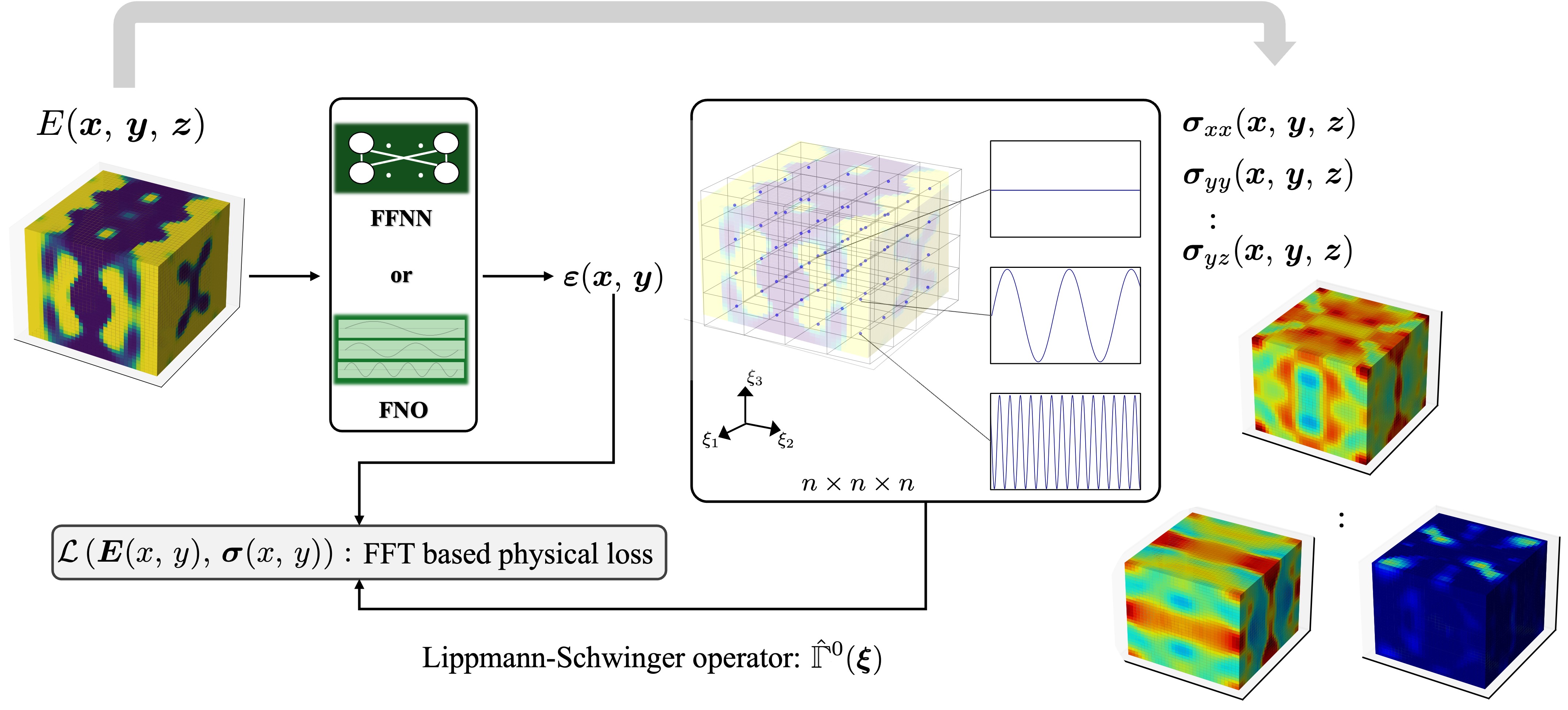}
  \caption{The SPiFOL framework (small deformation setup) maps heterogeneity maps from microstructure images to strain fields using either FFNN or FNO. Material models convert the strain to stress fields. The Lippmann-Schwinger operator is built based on the fixed finite output space and FFT-based physical loss is defined based on this operator.}  
\label{fig:spifol_arch}
\end{figure}

In this section, we present the SPiFOL architecture, which incorporates three key components: a standard multi-layer perceptron (MLP), a modified MLP inspired by \cite{wang2022improved}, and the Fourier Neural Operator (FNO) architecture \cite{li2020fourier}. A detailed schematic of the SPiFOL architecture for small deformations setup is shown in Fig.~\ref{fig:spifol_arch}. The SPiFOL framework is designed to map microstructural topologies to their corresponding mechanical responses (strains) in the context of small deformations. \\
In the context of finite deformations, this work focuses on a specific microstructure, where SPiFOL is designed to map various macroscopic deformation gradients applied to a microstructure to the corresponding full-field deformation gradient solutions.
The output fields predicted by the SPiFOL models (strains in small deformation setup and fluctuation parts of deformation gradient for the finite deformation case) are then converted to stress fields using the material constitutive laws specific to each phase within the microstructure. The Lippmann-Schwinger operator is then constructed based on the desired output resolution using Fourier shape functions, as shown on the right side of Fig.~\ref{fig:spifol_arch}. 
By leveraging gradients and physical constraints in Fourier space, the computational cost is significantly reduced. This approach eliminates the need for additional tools, such as AD or FEM, which can substantially increase training overhead. The equilibrium PDEs in Fourier space are efficiently established through a single multiplication of the Lippmann-Schwinger operator with the polarization stress, as outlined in Eqs.~\eqref{eq:8}. The following section provides a comprehensive discussion of the neural network architectures employed in this study.\\ 
\textbf{Remark 1:} Fourier space shape functions, used to compute gradients in Fourier space, can be constructed at multiple resolutions if sample data is available at those levels. This enables the formulation of physical constraints simultaneously across different resolutions, enhancing the model’s flexibility in multiscale applications.\\ 
\textbf{Remark 2:} The selection of shape functions in Fourier space is arbitrary; higher-order terms can be incorporated without increasing the computational cost of training. This flexibility allows for improved approximation capabilities without compromising efficiency.\\ 
\textbf{Remark 3:} The SPiFOL framework builds upon traditional FFT solvers, which inherently rely on rectangular, periodic microstructures and enforce periodic boundary conditions. While these constraints may appear restrictive, they are well-suited to multiscale modeling, where the assumption of periodicity effectively captures representative behavior across different scales.
\subsection{Networks architectures}
To provide a clear understanding of the network architectures, we will delve deeper into the structures of the FFNN and FNO architectures to illustrate a detailed explanation of how inputs are processed by each architecture. 

\subsubsection{FFNN models}
In this study, we employ two distinct neural network architectures.
The first architecture, referred to as MLP, is a vanilla FFNN without any encoder layers to map the input to the desired solution functions. The output of the $l_{th}$ layer of MLP is computed by 
\begin{equation}
\begin{aligned}
\label{eq:mlp}
z_m^l\,=\,act\left( 
\sum_{n=1}^{N_l} w_{mn}^l\,z_n^{l-1}\,+\,b_m^l \right),~~~ l =1,\, ..., ..., L-1.
\end{aligned}
\end{equation}
In Eq.\,\eqref{eq:mlp}, \textit{act} denotes the activation function, $m$ is the $m$-th component of $z$, and $l$ is the number of the $l$-th layer. $w_{mn}$ and $b_m$ are the corresponding weights and biases, respectively. $L$ is the total number of network layers and the last layer is a linear layer with no activation function. 

In addition to the MLP architecture, the modified MLP architecture is employed to further enhance the accuracy of the results. 
Building on the work of \cite{wang2022improved}, two encoders, $\bm{U}$ and $\bm{V}$, are introduced to enhance the network's capability, enabling the subsequent layers to better retain and recall the input function.
These encoders bring the input space into the feature space, which is used in each hidden layer by point-wise multiplication. These encoders are computed as 
\begin{equation}
\begin{aligned}
\label{eq:encoderU_V}
\bm{U}\,=\,\sum_{n=1}^{N_0} U^1_{mn}z^0_{n} + b_m^1, \quad \bm{V}\,=\,\sum_{n=1}^{N_0} V^1_{mn}z^0_{n} + b_m^2. 
\end{aligned}
\end{equation}
In Eq.\,\eqref{eq:encoderU_V}, $U^1_{mn}$ represents the weight connecting the $n$-th input to the $m$-th neuron in the encoder layer, where $\bm{z^0}$ denotes the input function, and $\bm{b^1}$ represents the biases. Similarly, $V^1_{mn}$ and $\bm{b^2}$ correspond to the weights and biases for the second encoder. Finally, the inputs of each of the subsequent layers are calculated by these encoders as
\begin{equation}
\begin{aligned}
\label{eq:znl_modified}
\bm{z}^{l}\,=\,\bm{z}^{l}\cdot\bm{U}\,+\,\bm{z}^{l}\cdot(1-\bm{V}).
\end{aligned}
\end{equation}

By incorporating Eq.\,\eqref{eq:encoderU_V}, and using $\bm{z}^{l-1}$ from Eq.\,\eqref{eq:znl_modified}, the modified MLP architecture is constructed.
\subsubsection{FNO model}
FNOs are excellent candidates for mapping microstructures to their corresponding stress fields due to the nature of the problem and the periodicity of microstructures. Unlike traditional CNNs that use local kernels, FNOs perform continuous convolution over the entire domain. This global convolution approach is inspired by the kernel formulation of solutions to linear PDEs using Green's functions \cite{li2020fourier, kovachki2023neural}.

Furthermore, FNOs exhibit resolution invariance, meaning the same FNO model can be applied to perform mappings even when the initial discretization is refined or coarsened. This capability is often referred to as zero-shot super-resolution (ZSSR). To leverage this property, in addition to the standard input—the microstructure map—a fixed input domain is provided in each dataset. Consequently, each dataset encodes spatial coordinates, \(\bm{x}, \bm{y}, \bm{z}\), along with an additional input channel representing the heterogeneity map.

In the FNO architecture within the SPiFOL framework, an initial dense layer P embeds the input function at each spatial point into a higher-dimensional latent space, denoted as $LS^0$, see Fig.\,\ref{fig:arch_fno}. This projection provides the necessary number of channels for the subsequent $L$ Fourier layers. Each Fourier layer transforms the input into the frequency domain using the Fourier transform $\mathcal{F}$, $R$ retains only the first few prescribed modes, and truncates the higher frequencies. A convolution is applied in this truncated Fourier space, and the result is brought back to the spatial domain using the inverse Fourier transform $\mathcal{F}^{-1}$. Finally, the output is passed through a non-linear activation function. Here, the Gaussian error linear unit ($\textit{Gelu}$) is defined as 
\begin{equation}
\begin{aligned}
\label{eq:gelu}
GELU (x) \approx 0.5x \left(1 + \tanh\left[\sqrt{\frac{2}{\pi}} \left(x + 0.044715x^3\right)\right]\right),
\end{aligned}
\end{equation}
Finally, the last dense network is used to project the outputs back to the target dimensionality corresponding to the strain fields at each point (\(\varepsilon_{xx}(x, y, z)\), \(\varepsilon_{yy}(x, y, z)\), \(\varepsilon_{zz}(x, y, z)\), \(\varepsilon_{xy}(x, y, z)\), \(\varepsilon_{xz}(x, y, z)\), and \(\varepsilon_{yz}(x, y, z)\)).
 The output of the $l+1$ layer of the Fourier layer is computed by
\begin{equation}
\begin{aligned}
\label{eq:FNO}
\bm{z}^{l+1}\,=Gelu\left(\mathcal{F}^{-1}(R \cdot \mathcal{F}(\bm{z}^{l}))\,+\, \bm{W}_l\cdot  \bm{z}^{l} + \bm{b}_l \right).
\end{aligned}
\end{equation}
In Eq.\,\eqref{eq:FNO}, $\bm{b}_l$ is the bias, $\bm{W}_l$ is the weight matrix, and $\bm{z}^{l}$ refers to the output of the previous Fourier layer or the input projected into $LS^0$. Thus, the weights and biases are designed to preserve the shape of the inputs. $R$ can be interpreted as a weight tensor that truncates higher modes in Fourier space (the same number of modes is utilized within each Fourier layer).
In this paper, isotropic elasticity is considered, where the material properties of each phase are characterized by two parameters. To simplify the input space, we vary only one of these parameters, Young's modulus 
$E$, across different phases. As a result, the ratio of the Lamé constants within each phase remains consistent when compared to other individual phases.

To identify each material phase, the microstructure image is pixelated to the desired resolution and used as an input layer for SPiFOL, as shown in Fig.\,\ref{fig:spifol_arch}. SPiFOL maps the input to the output field of interest, e.g., the strain fields $\bm{\varepsilon}$, in a physically informed manner (unsupervised learning). 

To enforce physical constraints, the balance of linear momentum in Fourier space based on the macroscopic strain, see Eq.\,\eqref{eq:8}, the corresponding Lippmann-Schwinger operator in Fourier space, see Eq.\,\eqref{eq:LippmanSch.Oper}, is defined prior to the training process and remains constant for the given discretization. These operators are independent of the microstructures' topologies. Similarly, see Eqs.\,\eqref{eq:F_update} and \ref{eq:LippmanSch.OperF} for the case of finite deformation.
\subsection{Constructing a physics-informed loss function}
In this work, three different input-output mapping methods are used: the standard MLP, a modified MLP, and the FNO architectures.
\subsubsection{Small deformation setup}
 In the case of small deformation, aforementioned networks map different microstructures to their corresponding strain fields ($\varepsilon^n_{N_{ij}}$, where $i,j \in {x, y, z}$).  
The loss function is formulated based on the output and the computed Lippmann-Schwinger operator, following the fixed-point scheme used in FFT-based mechanical solvers. The final loss function is computed as the mean squared error (MSE) of Eq.\,\eqref{eq:8}, with the right-hand side rearranged to the left-hand side. Due to the varying scales of the values in the loss function, resulting from the applied macroscopic strain, a weighting scheme is employed to normalize the individual components of the strain tensor to a comparable scale. This normalization ensures balanced contributions from each component, which improves the model’s accuracy. Further details of the training process will be discussed in later sections. The final loss is formulated by expressing Eq.\,\eqref{eq:8} in index notation for each individual voxel, $n$, as 
\begin{equation}
\begin{aligned}
\label{eq:L_eps_xi}
L_{ij}^n\,&=\, \,-\bar{\varepsilon}_{ij}\,+\,\varepsilon_{ij}(\bm{x}_n)\, +\, \mathcal{F}^{-1}\left\{
\hat{\Gamma}^{0}_{ijkl}(\bm{\xi}):\mathcal{F}\{\tau_{kl}(\bm{x}_n)\}
\right\},\quad \text{with} \quad \\\tau_{kl}(\bm{x}_n)\,&=\,\left(\mathcal{C}_{klmn}(\bm{x}_n)\,-\,\mathcal{C}^0_{klmn}\right):\varepsilon_{mn}(\bm{x}_n).
\end{aligned}
\end{equation}
In Eq.\,\eqref{eq:L_eps_xi}, \(\bm{x}_n\) represents the voxel $n$ corresponding spatial coordinates.
The indices \(i,\, j,\, k,\, l \in \{1,2,3\}\) refer to tensor components associated with the \(x\)-, \(y\)-, and \(z\)-directions. 
The total loss is calculated by first converting different components of the strain tensor to a comparable scale and then calculating the mean squared error of all weighted components as
\begin{equation}
\begin{aligned}
\label{eq:total_loss}
&\mathcal{L} =\, 
w_1\,\underbrace{\dfrac{1}{N_v}\sum_{n=1}^{N_v}(L_{11}^n)^2}_{\varepsilon_{xx}} 
\,+ \,w_2\,\underbrace{\dfrac{1}{N_v}\sum_{n=1}^{N_v}(L_{22}^n)^2}_{\varepsilon_{yy}} 
\,+\,w_3\,\underbrace{\dfrac{1}{N_v}\sum_{n=1}^{N_v}(L_{33}^n)^2}_{\varepsilon_{zz}} +
\\ &w_4\,\underbrace{\dfrac{1}{N_v}\sum_{n=1}^{N_v}\left((L_{12}^n)^2
+(L_{21}^n)^2\right)}_{\varepsilon_{xy}+\varepsilon_{yx}} 
    \,+\, w_5\,\underbrace{\dfrac{1}{N_v}\sum_{n=1}^{N_v}\left((L_{13}^n)^2
\,+\,(L_{31}^n)^2\right)}_{\varepsilon_{xz}+\varepsilon_{zx}} 
\,+\, w_6\,\underbrace{\dfrac{1}{N_v}\sum_{n=1}^{N_v}\left((L_{23}^n)^2+(L_{32}^n)^2\right)}_{\varepsilon_{yz}+\varepsilon_{zy}},
\end{aligned}
\end{equation}
where $N_v$ stands for the total number of voxels.
In Eq.~\eqref{eq:total_loss}, the coefficients $w_1$ through $w_6$ represent the weighting factors associated with each strain component. These coefficients can be adjusted based on the macroscopic strain $\bar{\bm{\varepsilon}}$ or optimized using advanced methods such as neural tangent kernels \cite{jacot2018neural, WANG2022why}. For a 2D case, the strain components related to the $z$-direction are excluded from the loss function.
\subsubsection{Large deformation setup}
\begin{figure}[H] 
  \centering
  \includegraphics[width=0.8\linewidth]{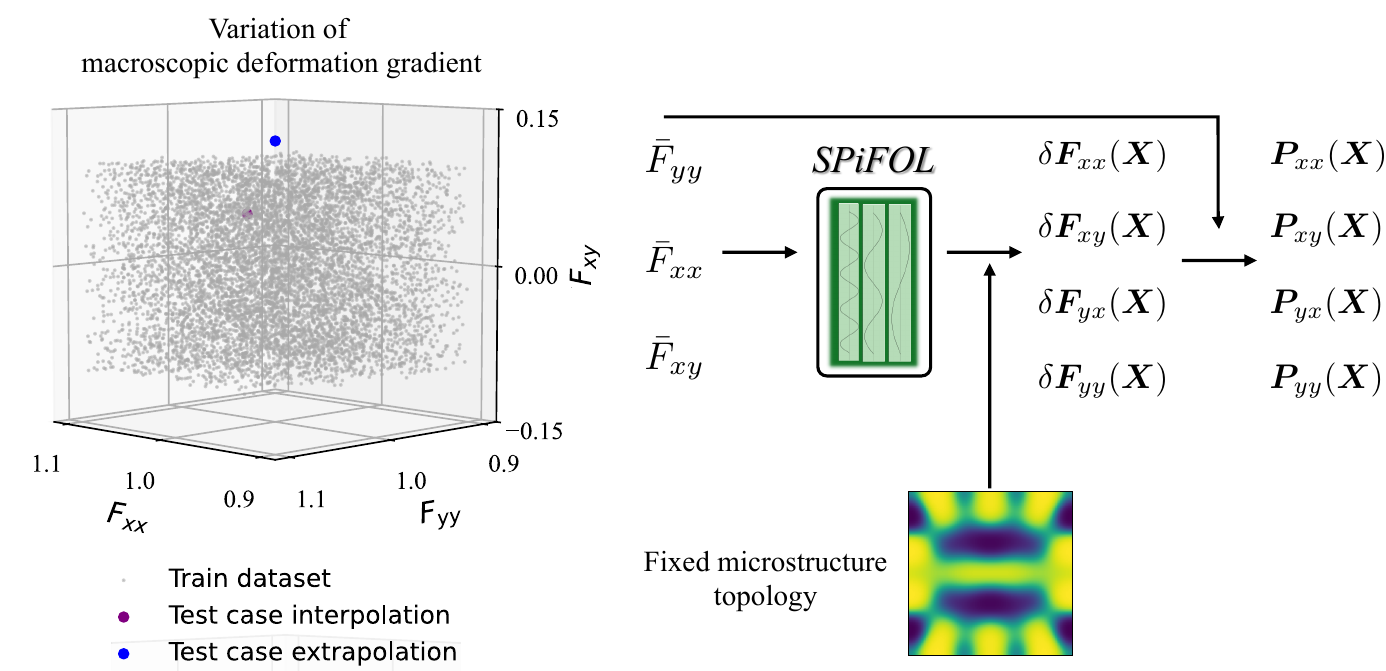}
  \caption{The SPiFOL finite deformation framework is trained on different macroscopic deformation gradient combinations applied to a fixed microstructure. It takes three macroscopic deformation gradient components as input and predicts the fluctuation part of the solution at each point, which is then used in the material model to calculate stresses.}
\label{fig:spifol_finite}
\end{figure}

In the context of finite deformation, SPiFOL is designed to learn the mapping between the input function space—such as microstructure topology or varying macroscopic boundary conditions, see Fig.\,\ref{fig:spifol_finite}, and the corresponding full-field solution for the fluctuation components of the deformation gradient. The total deformation gradient at each point is obtained by summing its fluctuation component, \(\delta\bm{F}(\bm{X})\), with the applied macroscopic deformation gradient, \(\bar{\bm{F}}\), as
\begin{equation}
\label{eq:total_f}
\begin{aligned}
\bm{F}(\bm{X}) = \bar{\bm{F}} + \delta\bm{F}(\bm{X}).
\end{aligned}
\end{equation}
In Eq.\,\eqref{eq:total_f}, $\bm{X}$ represents the position vector in the reference configuration.
Analogous to the small deformation case and Eq.\,\eqref{eq:L_eps_xi}, Eq.\,\eqref{eq:F_update} is also reformulated in index notation for voxel \(n\), as 
\begin{equation}
\begin{aligned}
\label{eq:loss_total_finite}
{L_F}_{ij}^n\,= -\bar{F}_{ij}\,+\,F_{ij}(\bm{X}_n)\, +\, \mathcal{F}^{-1} \left\{\hat{\Gamma}^{0, F}_{ijkl}(\bm{\xi}):\mathcal{F}\left\{{P}_{kl}(\bm{X}_n)\,-\,\mathcal{C}^0_{klmn}:{F}_{mn}(\bm{X}_n)
\right\}
\right\}.
\end{aligned}
\end{equation}
The first Piola-Kirchhoff stress tensor $\bm{P}$ is calculated using the material law and taking into account the deformation gradient $\bm{F}$.
 Finally, by substituting the components of Eq.~\eqref{eq:loss_total_finite} into Eq.~\eqref{eq:total_loss}, the total loss corresponding to the finite deformation case can be obtained. For the finite deformation case, no weighting factor is employed.
\section{Results}
\label{sec:results}
The SPiFOL framework, along with all the architectures proposed in this study, is implemented using \textit{JAX} \cite{jax2018}, which provides robust support for GPU-accelerated training.
All models are trained on an NVIDIA GeForce RTX 4090. For the FNO 2D small deformation setup, training on 8,000 samples over 30,000 iterations takes approximately 8 minutes. FFT solver time calculations are performed on an Apple M2 Pro CPU while SPiFOL model evaluation is done on the aforementioned GPU.
To optimize the network architecture and output normalization values, we use the Optuna package \cite{optuna_2019} within \textit{JAX}, which automates hyperparameter tuning based on a predefined objective function.
\begin{table}[htb]
\center
\caption{A list of hyperparameters for each network architecture is provided. The number of latent size, Fourier layers, Fourier modes, and activation function is the same across all FNO architectures.}  
\scriptsize
\begin{tabular}{|c|c|c|}
\hline
\rowcolor[HTML]{C0C0C0} 
architecture                   & hyperparameter                      & value    \\ \hline
                               & number of hidden layers                & 2        \\
                               & number of neurons in each hidden layer & 3500     \\
                               &\textit{act} & \textit{elu} \\
\multirow{-3}{*}{MLP}          & number of parameters                     & 51103072 \\ \hline
                               & number of hidden layers                & 2        \\
                               & num. of neurons in each hidden layer & 3500     \\
                               &\textit{act} & \textit{elu} \\                             \multirow{-3}{*}{modified MLP} & number of parameters                     & 58278072 \\ \hline
                               & latent size                         & 32       \\
                               & Fourier layers                      & 3        \\
                               & {Fourier modes}              & 16       \\
                               &\textit{act} & \textit{Gelu} \\          \multirow{-4}{*}{FNO 2D for small deformation setup}          & number of parameters                     & 1580771  \\ 
                               \hline 
                               FNO 2D  for finite deformation setup & number of parameters & 6299556 \\
                               \hline
                              FNO 3D& number of parameters & 6549282
                               \\
                               \hline
\end{tabular}
\label{tab:hyperparameters}
\end{table}
In this case, the objective function is the average total loss over the last $50$ iterations within $5000$ iterations of training. Table\,\ref{tab:hyperparameters} lists the optimal hyperparameters for each architecture. \ref{appendix:fno_study} also shows the studies on Fourier layers and Fourier modes of the FNO architecture. 

The size of the latent space, the number of Fourier levels, and the number of Fourier modes remain the same for all FNO models. However, the number of parameters varies due to differences in the number of input and output channels in each model. In the 2D case, where coordinates are defined in the $x$ and $y$ directions, the FNO small deformation setup has 1 additional input channel (phase value) and 3 output channels (strain components), while the FNO finite deformation setup has 3 additional input channels (macroscopic deformation gradient components) and 4 output channels, see Fig.\,\ref{fig:spifol_finite}. In the 3D case, which includes an additional $z$ coordinate, the FNO model takes the phase value as an additional input channel and maps it to 6 output channels. Note that in the 3D case, the total number of unknown Fourier modes increases significantly to construct a 3D FNO block, even though the number of modes in each spatial direction ($x$, $y$, and $z$) remains constant.

The \textit{ADAM} optimizer is utilized in this study. Due to the large number of samples and network parameters, our experiments with quasi-Newton optimizers, such as \textit{L-BFGS}, did not lead to any significant improvements, even when applied after \textit{ADAM}.

In the SPiFOL framework, the presence of complex numbers in the parameters of the FNO architecture, as well as their potential occurrence in the loss function due to the loss formulation, necessitates a modification of the standard \textit{ADAM} optimizer. This adjustment ensures proper computation of derivatives with respect to complex-valued parameters, see \cite{bassey2021survey}.
Additionally, the number of parameters for each model, along with the training and evaluation times, will be discussed in the subsequent sections.
For the small deformation setup (2D and 3D cases) the macroscopic strain tensor, denoted by $\bar{\bm{\varepsilon}}$, is fixed for all reported results, as
\begin{equation}
\begin{aligned}
\label{eq:macrostrain}
\bar{\bm{\varepsilon}}_{2D}\,=\,\begin{bmatrix} 0.05 & 0.0 \\ 0.0 & 0.0
\end{bmatrix}, \quad \text{and}\quad \bar{\bm{\varepsilon}}_{3D}\,=\,\begin{bmatrix} 0.05 & 0.0 & 0.0\\ 0.0 & 0.0 & 0.0 \\0.0 & 0.0 & 0.0.
\end{bmatrix}.
\end{aligned}
\end{equation}

For the material parameters, according to Eq.\,\eqref{eq:lame_phi}, the following material properties (Lamé constants) for both the stiffer and weaker phases are provided in the table. Table\,\ref{tab:phase_params}.
\begin{table}[H]
\centering
\caption{Material parameters for the different phases}  
\label{tab:phase_params}
\begin{footnotesize}
\begin{tabular}{ l l }
\hline
      & value/unit \\
\hline
Stiff phase ($\lambda_f$)  & $23.19$~GPa  \\
Stiff phase ($\mu_f$)  & $29.51$~GPa  \\
Soft phase ($\lambda_m$)  & $23.19/ r$~GPa  \\
Soft phase ($\mu_m$)  & $29.51/ r$~GPa  \\
\hline
\end{tabular}\\
$r$ denotes the phase contrast value.
\end{footnotesize}
\end{table} 
We use Eq.\,\eqref{eq:lame_phi} to calculate the material properties of the intermediate phase in Fourier-based microstructures. This study looks at four different phase ratio values $r$: $5$, $10$, $50$, and $100$.

This section is structured as follows: initially, we evaluate the performance of the top-performing SPiFOL networks, which include MLP, modified MLP, and FNO architectures, after training with $8000$ dual-phase samples.  Next, We compare the performance of the SPiFOL utilizing FNO architecture with the modified MLP one, which utilizes a reduced input space for finer resolution $64~\text{by}~64$. Subsequently, we compare the performance of the SPiFOL with an FNO architecture against data-driven FNO models and examine the ZSSR approach for multiple phase contrast ratio values.
The performance of SPiFOL extensions for 3D and finite elasticity problems is discussed at the end of this section.\\
\textbf{Remark 4:} With two fixed for maximum and minimum stiffness, the reference stiffness tensor remains constant for both dual-phase and Fourier-based samples, regardless of variations in phase topology. It depends solely on the phase contrast ratio. 
\subsection{Dual-phase microstructures} 
\begin{figure}[H]
  \centering
  \includegraphics[width=0.75\linewidth]{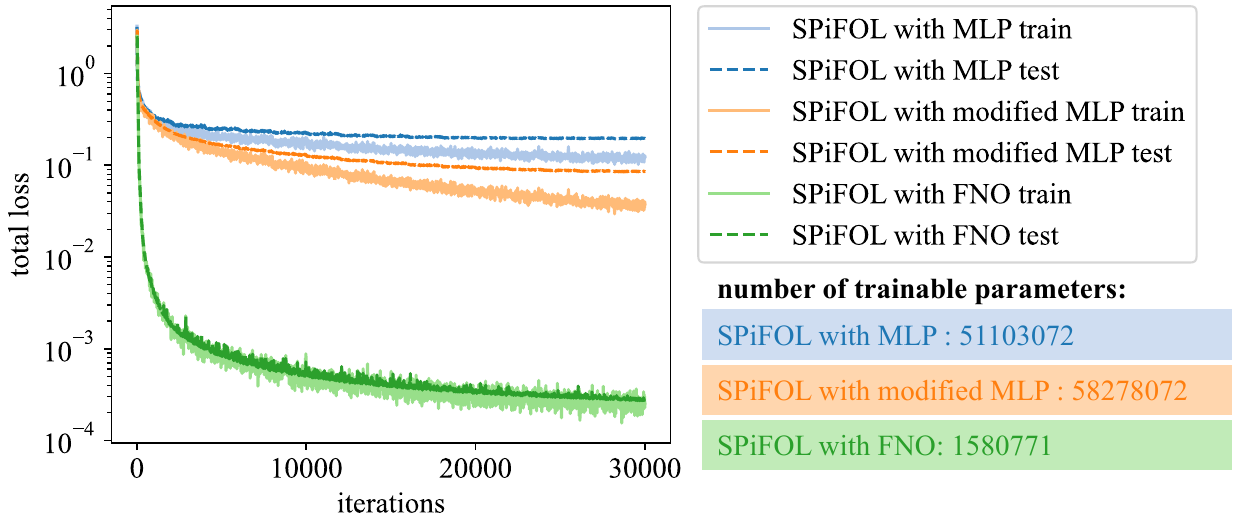}
  \caption{Loss decay for training and test cases for different network architectures, including MLP, modified MLP, and FNO, as well as the number of trainable parameters in each model.}
\label{fig:loss_train_test}
\end{figure}

The evolution of the loss function for the $r=5$ is shown on the left-hand side of Fig.\,\ref{fig:loss_train_test}. The training uses $8000$ samples of dual-phase microstructures depicted in \ref{2D_small_samples_dualphase}, with $20$ samples in each batch for all of the models. The weighting parameters introduced in Eq.\,\eqref{eq:total_loss} are selected to ensure that different components of the total loss are of the same order. This approach has been shown to yield better performance compared to scenarios where the orders of the loss components differ when it comes to PINNs, see \cite{WANG2022why, wang2022improved}.
The loss decays are depicted for various
 NN architectures considered in the SPiFOL framework, MLP, modified MLP, and FNO. The test loss is plotted for $100$ unseen microstructures. Fig.\,\ref{fig:loss_train_test} illustrates the train and test loss decay for all architectures as well as the number of trainable parameters in each case. 

\begin{figure}[H] 
  \centering
  \includegraphics[width=0.8\linewidth]{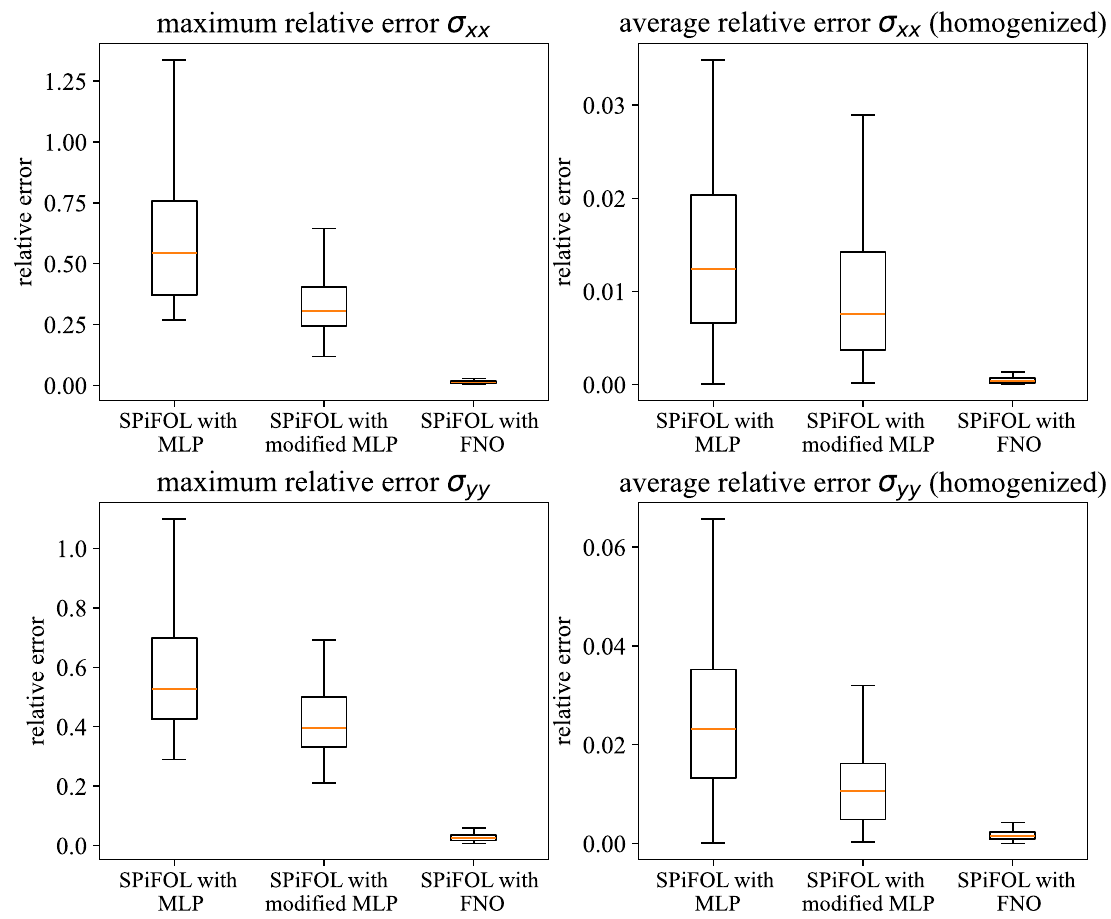}
  \caption{The relative maximum error (left-hand-side) and the relative average error (right-hand-side) for $\sigma_{xx}$ and $\sigma_{yy}$ for different SPiFOL architectures.}
\label{fig:avg_max_2_arch_error}
\end{figure}

 Fig.\,\ref{fig:avg_max_2_arch_error} presents the relative average error and maximum relative error across $100$ unseen cases. The SPiFOL model, which employs the FNO architecture, exhibits superior performance, with the relative maximum error remaining below $7\,\%$ for all cases. 

 The latter can reach a maximum of $120\%$ with MLP architectures. In terms of relative average stress error, all of the models exhibit a value below $5\,\%$. Modified MLP also shows better performance than MLP architecture. The average stress can be computed by $\bar{\bm{\sigma}}\,=\,\dfrac{1}{V}\int_V \bm{\sigma}(\bm{x},\,\bm{y})\,dV,$ where $V$ stands for the volume.

\begin{figure}[H] 
  \centering
  \includegraphics[width=1.0\linewidth]{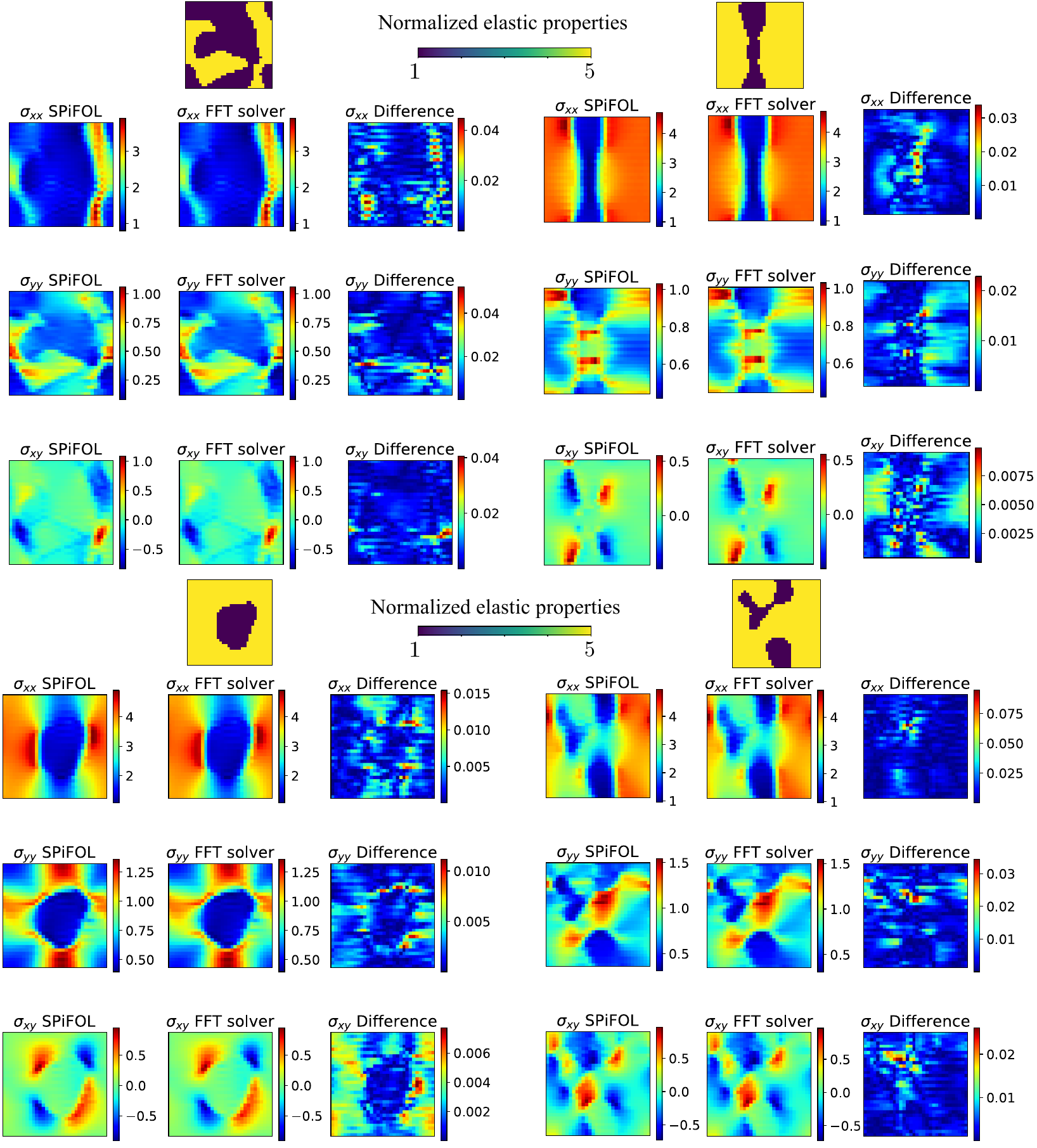}
  \caption{Prediction of stress fields for unseen microstructures using SPiFOL with an FNO architecture. The goal is to demonstrate the generalizability of SPiFOL in accurately predicting stress fields across various types of microstructures. All stresses have the unit of [$\text{GPa}$].}
\label{fig:results4test}
\end{figure}

As shown in Fig.\,\ref{fig:results4test}, SPiFOL demonstrates its applicability across a wide range of microstructures, even when the phase volume ratio varies significantly and multiple inclusions are present. 

\begin{figure}[H]
  \centering
  \includegraphics[width=0.99\linewidth]{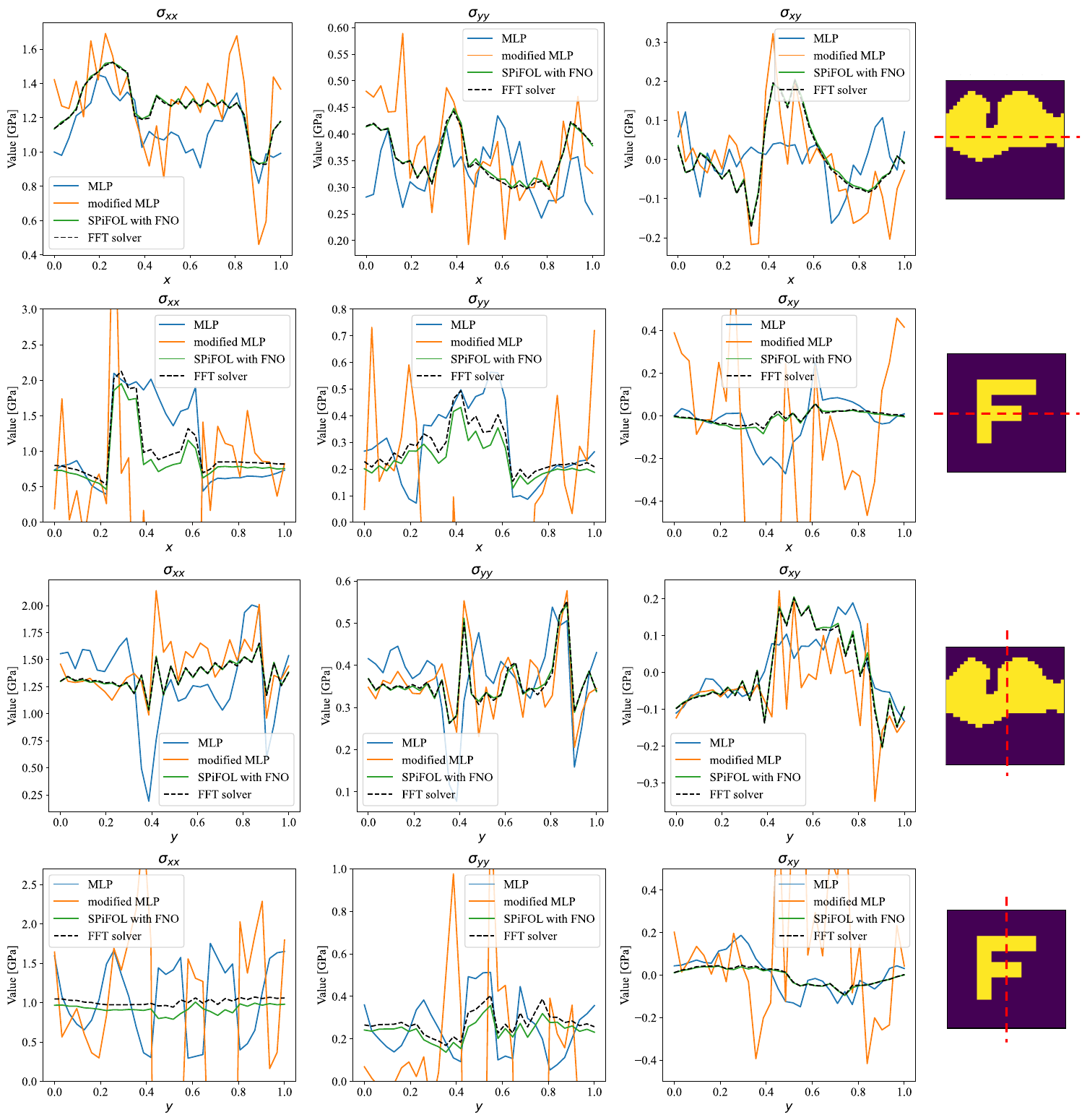}
  \caption{Comparison of the stress fields ($\sigma_{xx}$, $\sigma_{yy}$, $\sigma_{xy}$) for two random samples across two central sections, one in the $x$-direction and one in the $y$-direction. The results from the SPiFOL, utilizing different architectures, are compared against those from the FFT solver.}
\label{fig:sections}
\end{figure}

Fig.\,\ref{fig:sections} shows four random samples to demonstrate the accuracy of SPiFOL with different network architectures. The stress components are compared along two sections at the center of microstructures in the $x$ and $y$ directions. All of the architectures show acceptable performance and the predicted values are close to the reference solution obtained from the standard FFT solver. However, as Fig.\,\ref{fig:sections} depicts, the SPiFOL with MLP and modified MLP architecture sometimes overshoots the stress values significantly. The SPiFOL with FNO performs elegantly in predicting the right value of stress fields in the heterogeneous domain and even its fluctuation matches almost perfectly with the reference solution. In this study, Fourier space is employed in multiple instances, and a sample containing the letter 
$\bm{\text{F}}$ (stands for Fourier) is generated to assess the prediction accuracy of various models. The SPiFOL model, using the FNO architecture, continues to predict perfectly for this extrapolated sample. The MLP-based model provides reasonable predictions, while the modified MLP shows significant errors. This discrepancy in performance may be attributed to the large number of parameters in the modified MLP, potentially leading to overfitting.

\subsection{Fourier-based microstructures} 
In the last section, the performance of the SPiFOL framework for dual-phase microstructures was discussed. Building upon the motivation for architected materials, this section explores the application of the SPiFOL architecture to multiphase materials. These multiphase materials are generated using the functions outlined in section\,\ref{sec:fourier_samples}. The resolution is also increased for the output, and we tend to predict the stress at $64$ by $64$ resolution, whereas for dual-phase materials the resolution was $32$ by $32$. One of the major challenges in operator learning is increasing the resolution. As the resolution grows, the parametric space expands significantly, making it difficult to approximate the operator that maps the solution. This complexity arises from the need to capture finer details in the solution space, which demands more sophisticated methods to ensure accurate mapping. The two most effective SPiFOL architectures: SPiFOL with FNO and SPiFOL utilizing a parametric input space alongside a modified MLP architecture are employed. These models are trained using the $8000$ multi-phase samples.

\begin{figure}[htbp]
  \centering
  \includegraphics[width=0.99\linewidth]{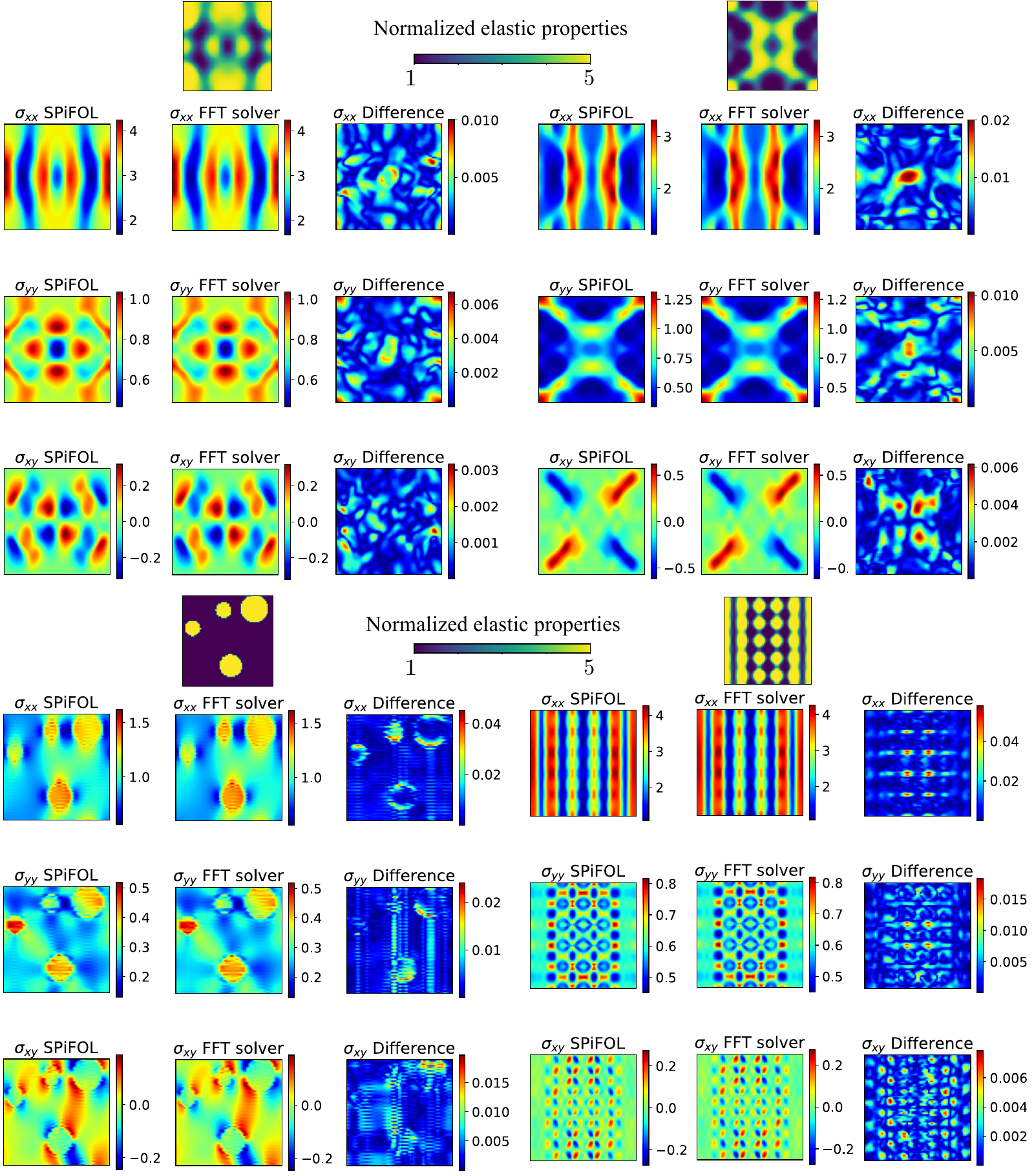}
  \caption{Prediction of stress fields for unseen multiphase microstructures using SPiFOL with an FNO architecture. The top are unseen microstructures generated using the same frequencies and normalized coefficients (interpolation cases). The bottom shows the extrapolation cases where a matrix with circular inclusions is depicted at the bottom left, and frequencies are changed at the bottom right to generate the new microstructure. All stresses have the unit of [$\text{GPa}$].}
\label{fig:results4test_multi_phase}
\end{figure}

To show the general applicability of trained SPiFOL, $4$ test cases are considered in Fig.\,\ref{fig:results4test_multi_phase}. The top two samples are unseen cases that are generated using the same frequencies and coefficients, see Eq.\,\eqref{eq:phi*} and Table\,\ref{tab:phase_params}. For the extrapolation cases, we considered two new samples: a matrix with multiple circular inclusions is shown, and a new set of frequencies is used to generate a new microstructure, as illustrated at the bottom left and right of Fig.\,\ref{fig:results4test_multi_phase}, respectively. 

\begin{figure}
  \centering
  \includegraphics[width=0.99\linewidth]{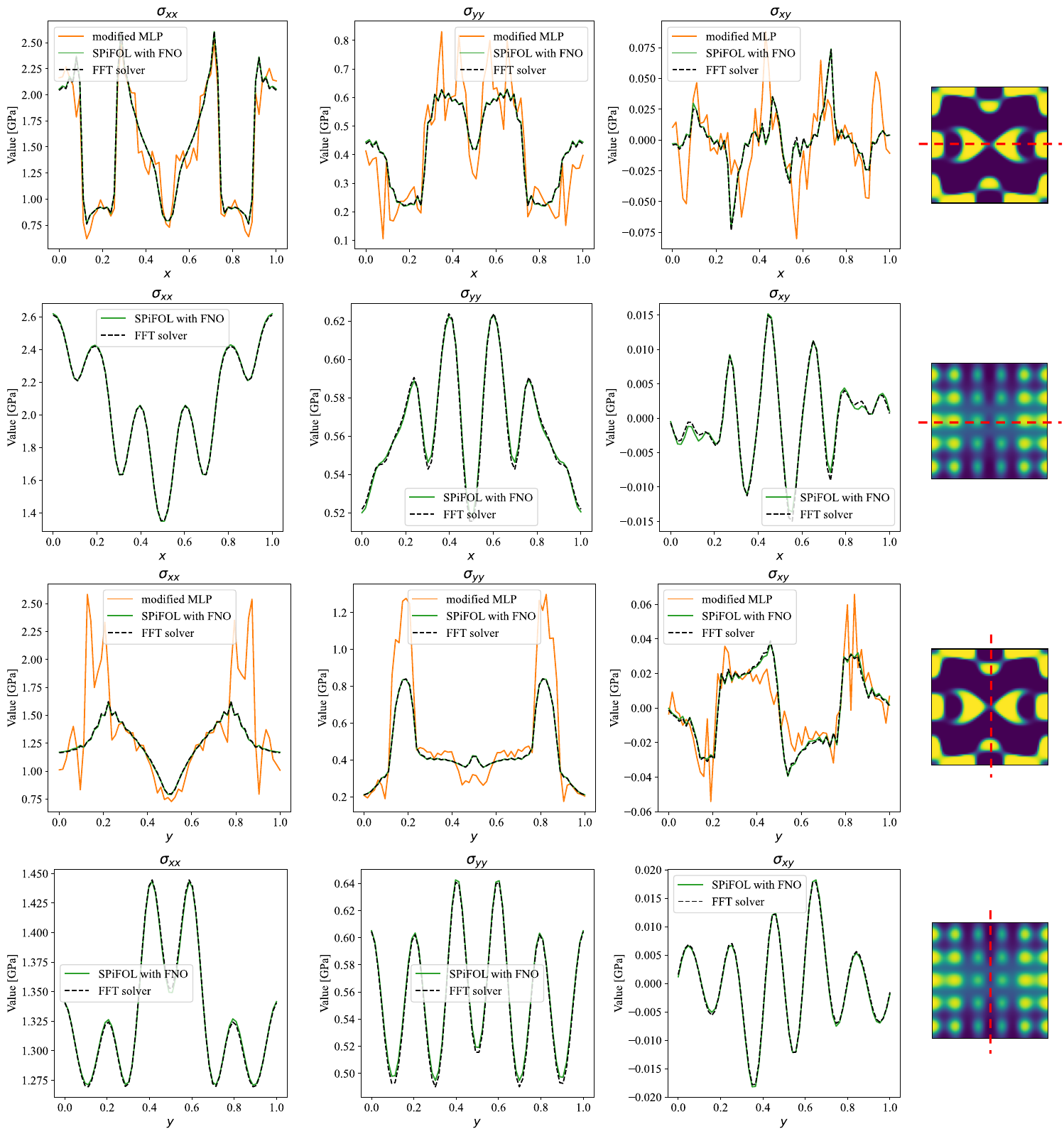}
  \caption{Prediction of stress fields for unseen multiphase microstructures using SPiFOL with an FNO architecture. The top is an unseen microstructure that was generated using the same frequencies and normalized coefficients (interpolation cases).  The bottom shows the extrapolation case where frequencies are changed completely to generate the new microstructures.}
\label{fig:section_comp_multi_phase}
\end{figure}
SPiFOL with FNO architecture does not lead to more than $1\,\%$ point-wise error
compared to the reference solution obtained from the FFT solver. For extrapolation cases, point-wise errors are still below $1.5\,\%$.

To make a more precise comparison, we evaluate two additional test cases: the first is generated using the same frequencies (top), and the second is produced using a new set of frequencies not included in the training data (bottom), as shown in Fig.\,\ref{fig:section_comp_multi_phase}. The SPiFOL model with FNO architecture is compared against the FFT solver and an alternative SPiFOL model with a modified MLP architecture, which leverages parametric input space compared through two cross-sections taken at the midpoint of the microstructures along the $x$ and $y$ directions. The SPiFOL model with the FNO architecture closely matches the solution obtained from the FFT solver. While the modified MLP model captures the overall patterns, it exhibits significant errors in peak regions, particularly for the extrapolation case, as illustrated in Fig.\,\ref{fig:section_comp_multi_phase} second and fourth rows. It is important to note that in the third test case, due to the parametric input space and the distinct set of frequencies used, the modified MLP architecture is no longer applicable, as the input space differs.
\subsection{Comparison of SPiFOL with FNO data-driven Models}
\begin{figure}[htb] 
  \centering
  \includegraphics[width=0.75\linewidth]{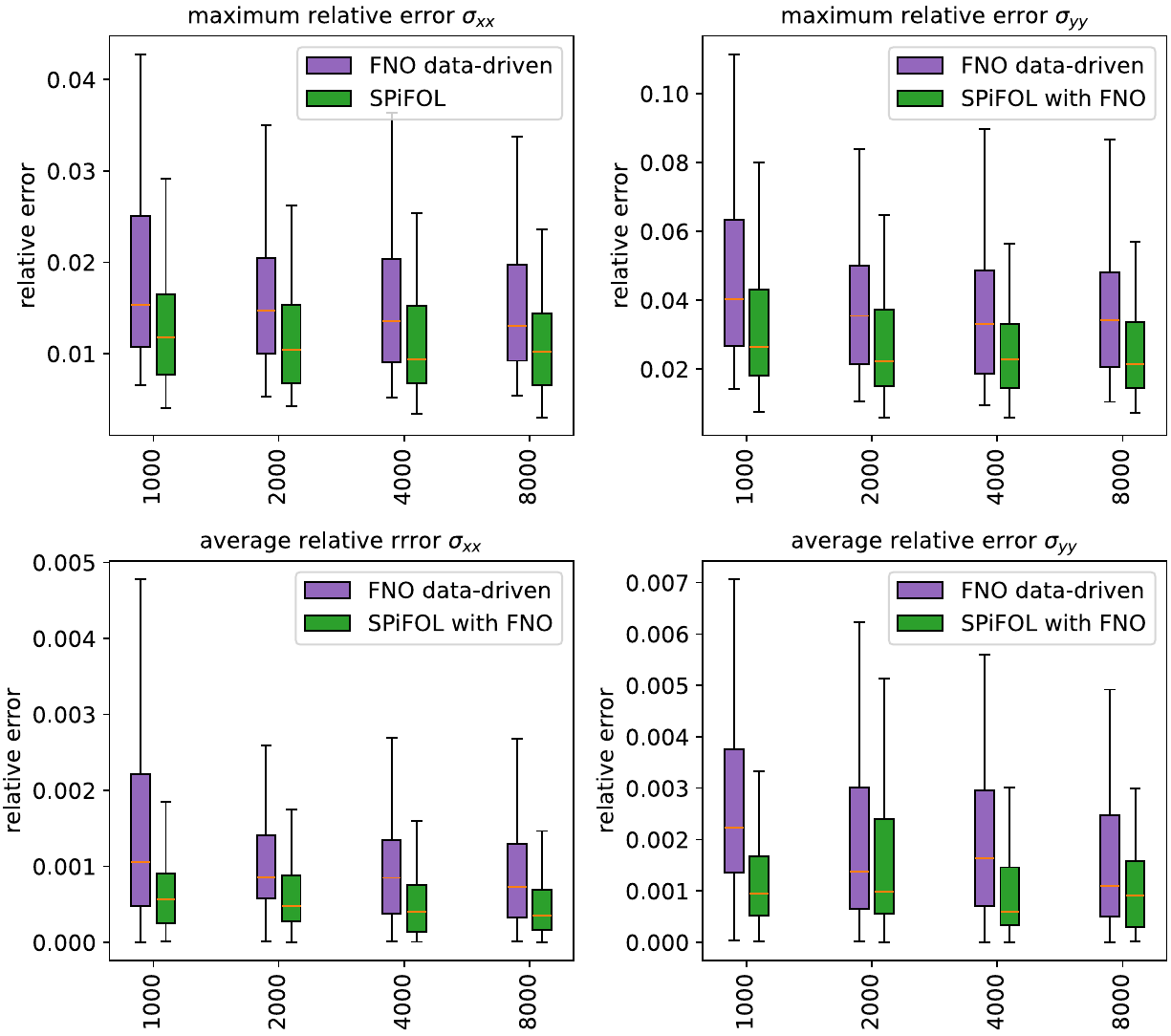}
  \caption{Comparison of maximum (top) and average (bottom) relative errors for $\sigma_{xx}$ (left) and $\sigma_{yy}$ (right) for data-driven FNO and SPiFOL with FNO. The relative errors are plotted against the number of samples the network has been trained on.}
\label{fig:fnodd_spifol_comp}
\end{figure}

In this section, we utilize the same FNO architecture, training the model in a fully data-driven manner using a dataset generated by the FFT solver.
To further highlight the effectiveness of these models, we evaluate their error performance on 100 unseen test cases. The models were trained using datasets of varying sizes: $1000$, $2000$, $4000$, and $8000$ dual-phase samples. For each model, we assess both the average relative error and the maximum relative error, including cases that require extrapolation.
The results show that increasing the number of training samples leads to a reduction in both the average and maximum relative errors. As illustrated in Fig.\,\ref{fig:fnodd_spifol_comp}, the maximum relative error for $\sigma_{xx}$ decreases from $0.04$ to $0.02$ as the training dataset size increases. However, the errors for $\sigma_{yy}$ are higher, likely because the absolute values of $\sigma_{yy}$ are smaller in comparison to $\sigma_{xx}$.

The data-driven FNO model shares the same architecture as the SPiFOL model with FNO, with both utilizing $16$ Fourier modes in each Fourier block. For additional details, please refer to \ref{appendix:fno_study}.
The SPiFOL model outperforms the data-driven FNO, reducing both the average and maximum relative errors by half, see Fig.\,\ref{fig:fnodd_spifol_comp}. Therefore, including physical equations in training in SPiFOL enhances the network prediction. \citet{li2024physics} also show the superior performance of adding physical constraints to the loss function in addition to data loss for Kolmogorov flows in which they use function-wise differentiation which is the explicit form of automatic differentiation. 
The latter is also achieved without increasing the training time, thanks to the  SPiFOL framework. The training and prediction times are compared in section\,\ref{sec:compute_time}.
\subsection{Zero-shot super-resolution (ZSSR) and phase contrast studies}
This subsection aims to evaluate the performance of SPiFOL across different phase contrast ratios. Leveraging the FNO framework, SPiFOL with FNO can predict responses at different resolutions, which is a key focus here. Specifically, the emphasis is on Fourier-based samples, as they allow for evaluation at different resolutions with having parametric microstructure details (see section\,\ref{sec:fourier_samples}). SPiFOL, utilizing the FNO architecture and the data-driven FNO models, is trained on $8000$ samples for four different phase contrast ratios: $5$, $10$, $50$, and $100$. 

\begin{figure}[H]
  \centering
  \includegraphics[width=1.00\linewidth]{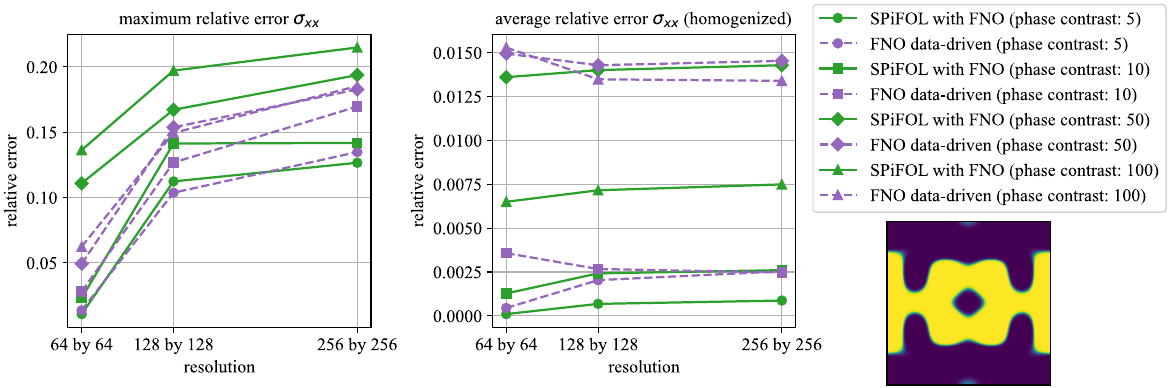}
  \caption{The effect of phase contrast on the maximum (left) and average (right) relative errors for different resolutions for SPiFOL with FNO and FNO data-driven models. For higher resolutions than the trained one (64 by 64), we utilize the ZSSR power of FNO to compute the solutions for a sample depicted on the bottom right side.}
\label{fig:phase_resolution}
\end{figure}

Fig.\,\ref{fig:phase_resolution} demonstrates that increasing the phase contrast leads to a significant rise in both maximum and average relative errors for both the SPiFOL with FNO and the FNO data-driven models. Notably, the maximum error increases substantially at higher resolutions, especially beyond the resolution at which the models were originally trained. However, the average relative error remains consistent across different resolutions for each phase contrast. This consistency highlights the strong potential of the proposed methodologies in effectively homogenizing the microstructural response across varying resolutions.
 The maximum error of SPiFOL goes beyond the data-driven FNO model in the case of high phase contrast ratios $50$ and $100$. The latter can highlight the shortcomings of the basic scheme in FFT-based approaches to the treatment of Gibbs oscillations \cite{schneider2021review, danesh2023challenges}.

\begin{figure}[htb] 
  \centering
  \includegraphics[width=0.9\linewidth]{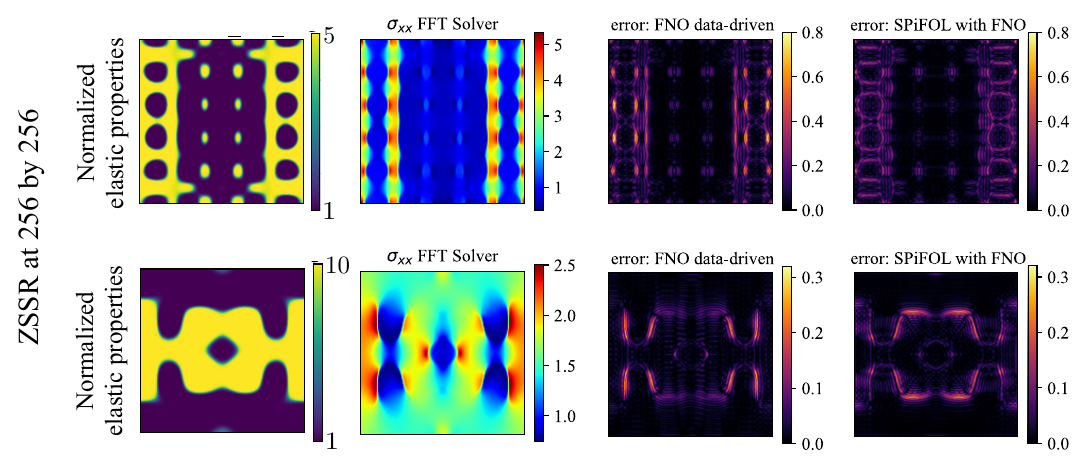}
  \caption{Comparison of ZSSR feature in SPiFOL with FNO and data-driven FNO at 256 by 256 resolution for two unseen cases with the phase contrast of $10$.}
\label{fig:fnodd_spifol_zssr_comp}
\end{figure}

The performance of ZSSR for SPiFOL and data-driven FNO is illustrated in Fig.\,\ref{fig:fnodd_spifol_zssr_comp} for the phase contrast of $10$. The maximum error increased by an order of magnitude, reaching around $10\,\%$ for both the data-driven FNO and the SPiFOL with FNO architecture. However, SPiFOL consistently demonstrates a lower maximum error than the data-driven FNO in both cases. The top case in Fig.\,\ref{fig:fnodd_spifol_zssr_comp}, highlights the differences between the two methods, demonstrating SPiFOL's superior performance in extrapolation scenarios due to its incorporation of physical constraints.

\begin{figure}[H]
  \centering
  \includegraphics[width=1.02\linewidth]{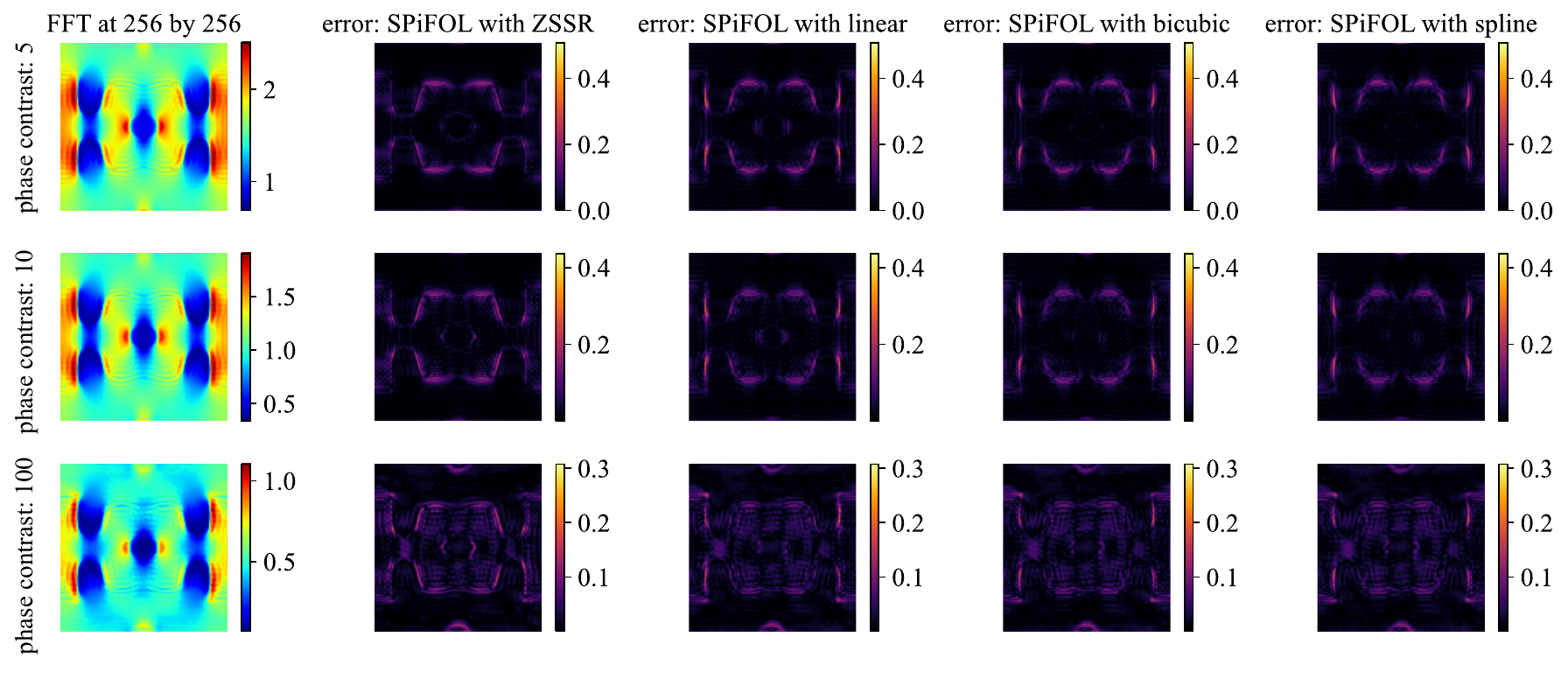}
  \caption{ZSSR feature of FNO (second column) is compared with linear interpolation (third column), bicubic interpolation (fourth column) as well as spline interpolation (fifth column) for different phase contact values.}
\label{fig:interpolations}
\end{figure}

To provide a more comprehensive analysis, we also applied two interpolation methods - bicubic and spline interpolation - to address the ZSSR capability of FNO. Specifically, the SPiFOL model, trained and evaluated at a resolution of $64$ by $64$, was interpolated at a resolution of $256$ by $256$ by the mentioned interpolating methods. Thanks to the ZSSR capability, the SPiFOL with FNO trained at the same evaluation is directly evaluated at a finer resolution in Fig.\,\ref{fig:interpolations}. 
The ZSSR outperformed linear, cubic, and spline interpolation methods in terms of the maximum error. The latter has also been addressed in \cite{sinha2024effectiveness}. Interestingly, as illustrated in Fig.\,\ref{fig:interpolations}, the locations of the maximum error for SPiFOL differ from those produced by interpolation methods, which tend to concentrate errors in regions with the highest values. In contrast, SPiFOL's maximum errors occur in mid-range values, which can be advantageous in cases where the maximum stress values are critical.
Additionally, as the phase contrast value increases from $5$ to $100$, the maximum error decreases. However, the error has become more widespread, appearing in more locations than before. For the $100$, the ZSSR error lies in the same order as other methodologies. 

\begin{figure}[H]
  \centering
  \includegraphics[width=0.99\linewidth]{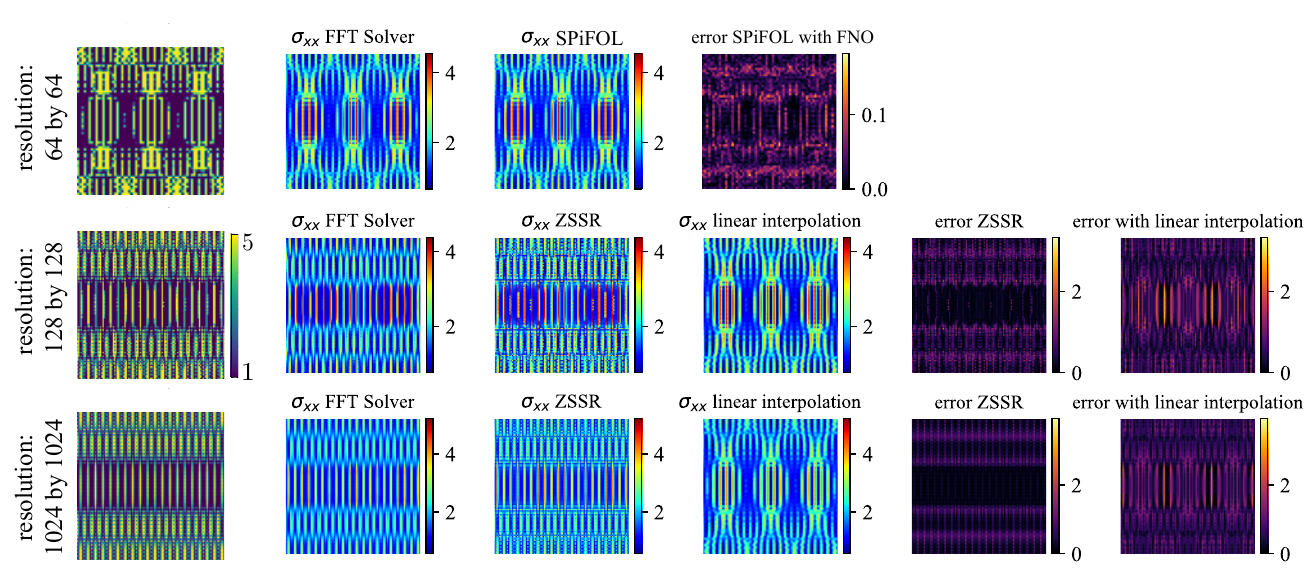}
  \caption{Comparison of ZSSR with linear interpolation for a high frequency generated sample. The SPiFOL is trained on the $64$ by $64$ resolution and evaluated on $128$ by $128$ and $1024$ by $1024$ microstructures. The linear interpolation results on $64$ by $64$ by using the linear interpolation to be evaluated in the higher resolutions.}
\label{fig:zssr1024}
\end{figure}

The interpolation techniques shown in Fig.\,\ref{fig:interpolations} demonstrate robust performance for the samples displayed in Fig.\,\ref{fig:fnodd_spifol_zssr_comp}. Since the lower-resolution samples are derived from lower frequencies, they do not exhibit significant information loss. To further evaluate the effectiveness of the ZSSR method and to allow for a comprehensive comparison, we used extremely high frequencies to generate multiphase samples. This approach is consistent with the methods described in section\,\ref{sec:fourier_samples}  as illustrated in Fig.\,\ref{fig:zssr1024}. 

\begin{figure}[htb]
  \centering
  \includegraphics[width=0.99\linewidth]{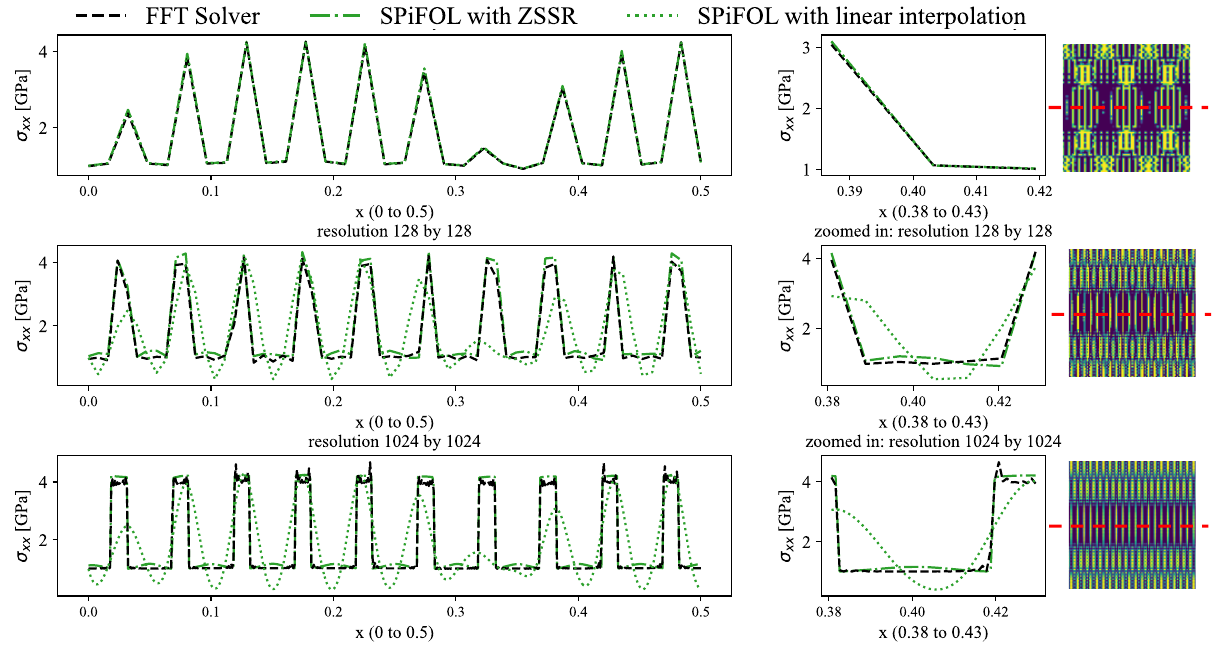}
  \caption{Comparison of ZSSR with linear interpolation for a high frequency generated sample. The SPiFOL is trained on the $64$ by $64$ resolution and evaluated on $128$ by $128$ and $1024$ by $1024$ microstructures. The linear interpolation results on $64$ by $64$ by using the linear interpolation to be evaluated in the higher resolutions.}
\label{fig:zssr_section}
\end{figure}

Although using ZSSR at higher resolutions increases the error, its superior performance and improved pattern recognition - achieved by analyzing the sample at a new resolution - are significantly better than those of simple interpolation techniques evaluated using results obtained from coarse samples.
To further show the effectiveness of ZSSR, we compare the results obtained from ZSSR and linear interpolation with those of the FFT solver along a cross-section made at the center of the samples in each resolution in Fig.\,\ref{fig:zssr_section}. These sections end at the midpoint (($x=0.5$) due to the symmetry of results, and on the right side of each case, a zoomed plot is made between $x=0.38$ and $x=0.43$. 

The obtained results from ZSSR can catch the peaks along the cross-section whereas the linear interpolation shows magnificent errors around the peaks and also middle points. However, since the SPiFOL with FNO architecture is trained on the coarse resolution of $64$ by $64$ resolution it cannot represent the fluctuations around the peak accurately, see zoomed plots in the right-hand side of Fig.\,\ref{fig:zssr_section}.

\subsection{Extension to 3D} 
In this section, the SPiFOL framework is extended to 3D problems. For the small strain setup, the network considers the 3D microstructure topology and predicts all strain components in 3D, including $\bm{\varepsilon}_{xx},\, \bm{\varepsilon}_{yy},\, \bm{\varepsilon}_{zz},\, \bm{\varepsilon}_{xy},\, \bm{\varepsilon}_{xz}, \, \bm{\varepsilon}_{yz}$. The Fourier blocks have been modified to accommodate the 3D setup. Training is performed on Fourier-based samples, using two additional frequencies for sample generation. The details of the sample generation process are described in \ref{appendix:3d_dataset}.
Similar to the previous section, training is performed on 8,000 samples, as shown in Fig.\,\ref{fig:3ddataset}, using the FNO architecture.
\begin{figure}[H]
  \centering
  \includegraphics[width=0.90\linewidth]{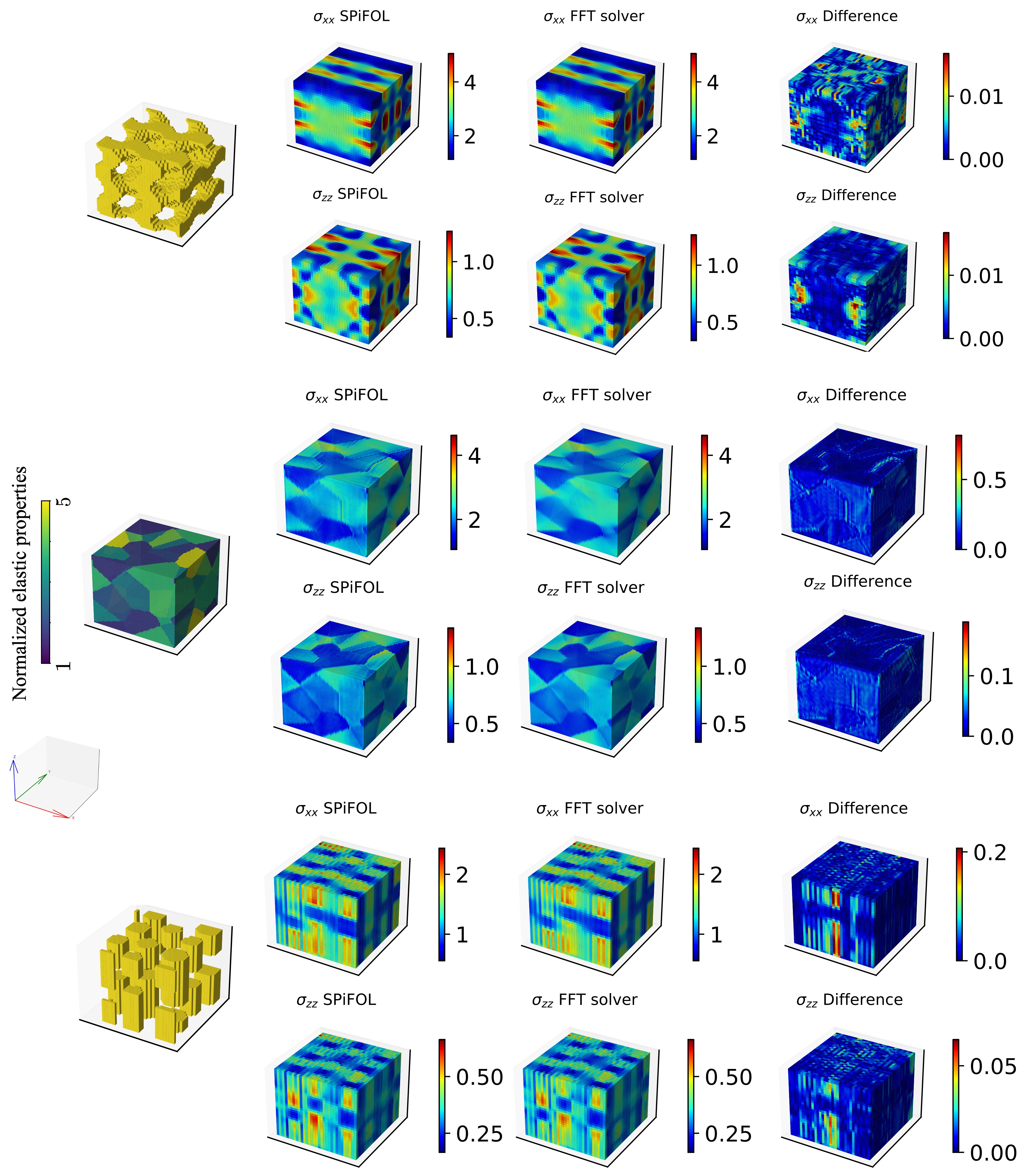}
  \caption{Predicted stress fields for three extrapolation cases. Top: TPMS-like structure; Middle: 3D polycrystalline material; Bottom: short fiber composite microstructure. Stress values are reported in [GPa]. For the TPMS-like and short fiber composite materials, only the strong phase is visualized.}
\label{fig:3d_results}
\end{figure}

The Fourier layer is specifically designed to ensure that the Fourier modes used in the network maintain symmetry. This design choice reduces the total number of network parameters and ensures that the learned solutions respect the inherent symmetries of the problem.
To evaluate the extrapolation capability of the network, we consider two unseen cases that deviate significantly from the sample distributions of polycrystalline microstructures and TPMS-like materials (dual-phase microstructure in which the strong phase is built like TPMS materials). The network predictions for these cases are presented along with the corresponding reference solutions obtained by the FFT solver. In addition, the difference between the SPiFOL model - incorporating the FNO architecture - and the reference solution is analyzed to evaluate the prediction accuracy.

For the extrapolation cases presented in Fig.\,\ref{fig:3d_results}, the error remains on the same order of magnitude as in the 2D cases. This consistency is attributed to the global kernel approximation in Fourier space, which enables SPiFOL, with its FNO-based architecture, to effectively learn the mapping between 2D or 3D microstructures and their corresponding mechanical responses. In the case of multiple cylindrical short fibers, the prediction error increases. This indicates that when the solution exhibits a higher spectral bias, the model must be trained on samples that are more representative and closer to the test cases. Additionally, increasing the neural network’s capacity becomes essential to capture the underlying complexity accurately.
\subsection{Extension to finite elasticity}
\label{sec:finite_elas} 
In the context of finite deformation, SPiFOL is trained to predict the mechanical response of a given microstructure, assuming the Saint-Venant material law for both phases. The network first estimates the deformation gradient, which is then used to compute the corresponding stress fields.  

To maintain the consistency in the output components of the network, the deformation gradient is rewritten in terms of its fluctuation component, \(\delta \bm{F}\). This redefinition is critical because the diagonal components of the deformation gradient are approximately 1.0, while the off-diagonal terms are close to 0.0, ensuring numerical stability and improved learning efficiency.  

The model is trained on a data set of 8,000 samples covering a wide range of macroscopic strain combinations in a 2D setup. The diagonal components, \( F_{xx} \) and \( F_{yy} \), vary in the range [0.9, 1.1], while the shear components of the deformation gradient vary between -0.1 and 0.1, see Fig.\,\ref{fig:spifol_finite} that shows the input space of SPiFOL for finite deformation. This extensive dataset allows the model to effectively generalize across different deformation scenarios.

The network considers the macroscopic deformation gradient \(\bar{\bm{F}}\) at the center of the microstructure and predicts the full-field deformation gradient components throughout the microstructure. The network inputs consist of the macroscopic deformation gradient components, represented as \(\begin{bmatrix} F_{xx} & F_{xy};& F_{yx} & F_{yy} \end{bmatrix}\). 
 The trained SPiFOL model can be used as a surrogate model for the homogenization of a specific microstructure subjected to varying deformations. 
To assess the predictive performance of the trained SPiFOL model, two test cases are analyzed, representing both interpolation and extrapolation scenarios. In the interpolation case, the macroscopic deformation gradient falls within the range of training values. In contrast, for the extrapolation case, the macroscopic deformation gradient exceeds the values encountered during training.

It is worth mentioning that to solve the same problem using the FFT solver, macroscopic loading is applied within 50 steps. Furthermore, for the extrapolation case, we observe that the fixed point scheme does not converge to the solution. Therefore, the Newton-based approach proposed in \cite{DEGEUS2017412} is employed. \\
\textbf{Remark 5.} By considering the symmetry in strains for the small deformation setup which reads the symmetry for stress tensor. For the finite deformation case, the means square error for $50$ different macroscopic deformation gradient cases are considered and the following error criterion is computed as
$(\bm{P}.\bm{F}^T -\bm{F} \cdot \bm{P}^T)^2/50$.
which is around $10^{-11}$, and shows the fulfillment of angular momentum.
\begin{figure}[H]
  \centering
  \includegraphics[width=0.99\linewidth]{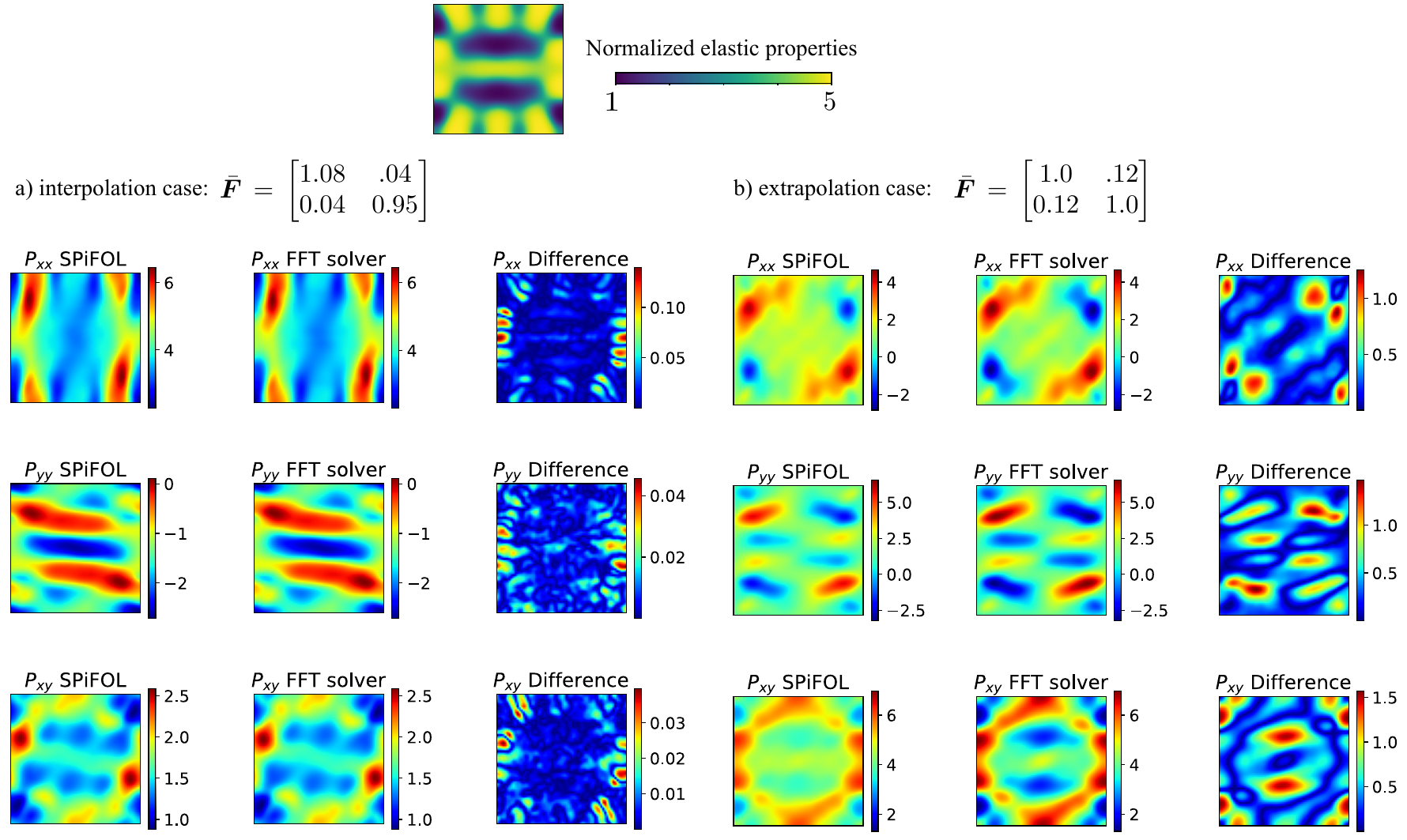}
  \caption{Prediction of the first Piola-Kirchhoff stresses:
(a) An unseen interpolation case, where the macroscopic deformation gradient components fall within the training range.
(b) An extrapolation case, where the deformation gradient component values extend beyond the range of the training data.}
\label{fig:finite_comp}
\end{figure}
\subsection{Discussions on computational cost}
\label{sec:compute_time} 
SPiFOL utilizes Fourier-based shape functions (frequencies) on a fixed finite discretization of the output space to compute gradients in Fourier space. The Lippmann-Schwinger operator is precomputed before training, and the PDE loss is defined by applying this operator to the network outputs. Consequently, the training time for SPiFOL is comparable to that of traditional data-driven FNOs.
Furthermore, the output data should not be randomly selected to create batches during each training iteration, which makes SPiFOL slightly faster than data-driven FNOs ($5~\,\%$ less computational cost for the cases of 2D small deformation). 
The training time on a GPU (NVIDIA GeForce RTX 4090) is approximately 24 minutes for the 2D case (small deformation setup with 8,000 samples), increasing to around 200 minutes for the 3D case. For the finite deformation scenario in a 2D setup, training takes roughly 40 minutes. Notably, all SPiFOL models employ a purely physics-informed methodology, eliminating the need for time-consuming ground truth data generation.
Once the network is trained, its evaluation to predict solutions is swift, taking only a fraction of a second thanks to the forward inference.

\begin{figure}[H]
  \centering
  \includegraphics[width=0.99\linewidth]{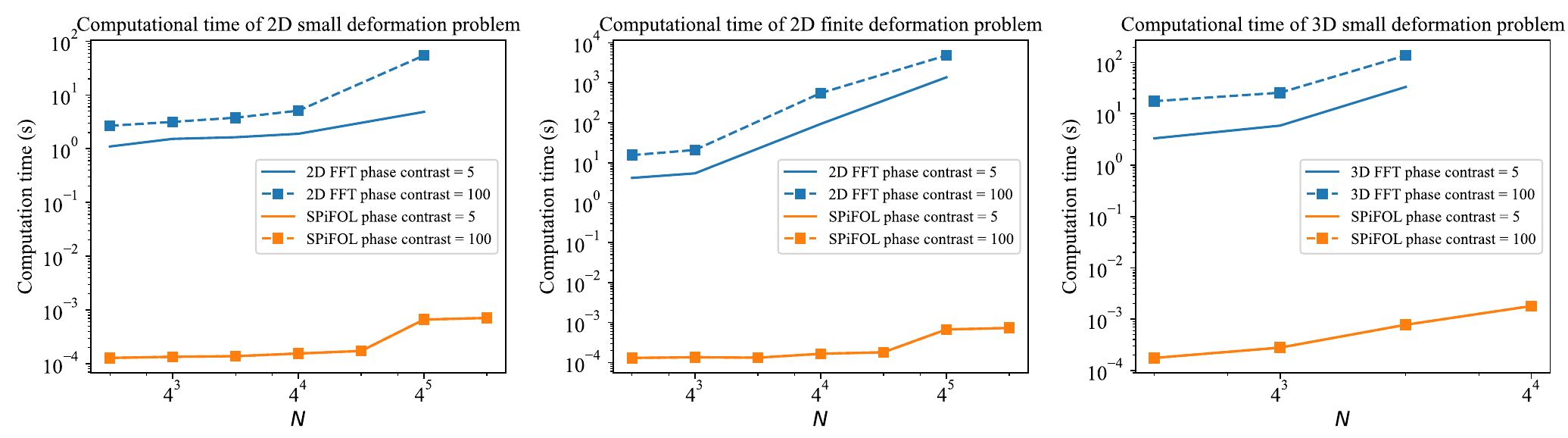}
  \caption{Comparison of computational time of SPiFOL with conventional FFT solvers for the phase contrast ratios of $5$ and $100$. Left: 2D small deformation setup with training time of 24 minutes. Middle: 2D finite deformation of the case with training time of 40 minutes. Right: 3D small deformation case with training time of 200 minutes. Note that training times are not included in the plots, and since all models are data-free, no time is required for data generation. $N$ stands for the number of grid points in each direction.} 
\label{fig:comp_time}
\end{figure}

The effectiveness of SPiFOL with FNO architecture is evaluated, and its computational time (only forward inference is considered for SPiFOL) is compared with conventional FFT solvers in Fig.\,\ref{fig:comp_time}. For the small deformation cases 2D and 3D, the FFT fixed point scheme is applied for two different phase contrast ratios. The values shown are the average of $20$ different microstructures.    

The SPiFOL is trained on the $64$ by $64$ resolution. Models utilizing the FNO architecture are trained with a latent size of 32, and the power of ZSSR is applied across different resolutions. Since the network evaluation depends solely on the network parameters and the employed resolution, the evaluation time remains constant for all phase contrast ratios.

For finite elasticity, the 2D SPiFOL model is evaluated against the Newton-based FFT approach proposed by \citet{DEGEUS2017412}. The calculation of the FFT solver and the evaluation of SPiFOL are performed on the CPU (Apple M2 Pro) and the mentioned GPU, respectively, to achieve the fastest computation time.
\\ 
\textbf{Remark 6:} For higher phase contrast ratios, more efficient FFT-based algorithms are available that can reduce the solver time to match the reported time for lower phase contrast values. However, the computation of the Lippmann-Schwinger operator will become more time-consuming in these cases, see \cite{kabel2014efficient}. 

\section{Conclusions and outlook}
In this work, we introduce a novel spectral physics-based operator learning method called Spectral Physics-Informed Finite Operator Learning (SPiFOL). This method is trained in a purely physics-informed manner, without relying on any ground truth data to build a general elasticity surrogate model. SPiFOL is designed to map 2D or 3D microstructure topologies to their corresponding strain fields by minimizing a physical loss function that represents mechanical equilibrium in Fourier space, under a given macroscopic strain. In the finite deformation case, SPiFOL is designed to map various macroscopic deformation gradients applied to a given microstructure to their corresponding full-field mechanical solutions.

The physical loss function is constructed using the finite operator learning methodology, where the output function is discretized on a fixed domain—in this case, in Fourier space. This approach avoids the need for automatic differentiation, which is commonly used when building physical loss functions. By employing the definition of the Lippmann-Schwinger operator in Fourier space, SPiFOL computes gradients with almost no additional computational cost. This results in training costs that are comparable to conventional data-driven operator learning methods, such as the FNO. Moreover, our results demonstrate that the accuracy of the proposed SPiFOL methodology surpasses that of purely data-driven FNOs. Specifically, the FNO architecture in SPiFOL achieves a maximum relative error of $2\,\%$ and a relative average error below $0.3\,\%$ for the same class of microstructures.

It is well known that operator learning models, including physics-informed variants such as SPiFOL, offer substantial computational speed-ups during inference (up to $200$ times, depending on the phase-contrast ratio). However, the fidelity of these models is sensitive to how representative the training sample set is of the target microstructure or loading conditions. To mitigate this, our training samples are generated to span a diverse set of microstructural features and boundary conditions. Moreover, the inclusion of physics-based loss terms improves the extrapolation capability. Still, for microstructures significantly different from the training distribution, some loss in accuracy is expected. To address this, the method can be further enhanced using meta-learning strategies, where additional adaptation steps during training help decode each new microstructure separately \cite{asl2025}. Another way to tackle this problem is to combine the current methodology with an attention-based mechanism to fine-tune the model for a specific class of microstructures, see \cite{xiao2023improved,calvello2024continuum}. 
In the case of a single microstructure, the on-the-fly SPiFOL model leads to a higher computational cost (about 8 times more iterations compared to conventional FFT solvers for reaching the same accuracy), depending on factors such as the phase contrast ratio, material nonlinearity, and network initialization (see~\ref{appendix:otf} for on-the-fly model results). We note that, unlike FFT solvers, neural network models may face challenges such as getting trapped in local minima because of their stochastic gradient optimization nature.

Additionally, by leveraging the ZSSR capabilities of FNOs, the network can predict responses at different resolutions. However, the accuracy tends to decrease when the network is applied to finer resolutions than those it was trained for. This was also addressed by previous works such as \cite{sinha2024effectiveness, tran2021factorized}. We also observe that SPiFOL leads to lower test loss values when it benefits from the FNO architecture, and both test and training losses converge to similar levels when the training samples are sufficiently diverse. This highlights the potential of SPiFOL as a robust operator learning technique that does not require labeled data.
While SPiFOL demonstrates remarkable accuracy and efficiency in capturing solutions to parametric PDEs, its applicability is inherently limited by the need to guarantee problem periodicity, restricting its use to specific problem types. Furthermore, incorporating physical constraints directly in Fourier space can be challenging and requires careful formulation to ensure consistency with the underlying physics. These limitations must be addressed when extending the method to more complex or non-periodic problems.

The methodology proposed in this work can be easily extended to other computational mechanics problems to minimize the governing PDEs with almost no additional computational effort. In addition, SPiFOL can be trained simultaneously at multiple resolutions. For example, in the case of the elasticity problem, building multiple Lippmann-Schwinger operators at different resolutions could further improve the accuracy of the ZSSR predictions.
Future work should focus on extending SPiFOL to handle highly nonlinear, path-dependent problems such as plasticity. This requires mapping multiple microstructure states to their subsequent configurations, as suggested in \cite{wang10428110}. In addition, the incorporation of the implicit FNO approach, as suggested in \cite{jiang2025implicit}, could further enhance the model's ability to capture complex path-dependent behavior.
In future works, we aim to leverage SPiFOL to develop surrogate models for phase transformation PDEs, such as the Allen-Cahn and diffusion equations, enabling operator learning without the need for data. 
Additionally, the high accuracy of the proposed methodology makes SPiFOL highly suitable for inverse problems and sensitivity analysis, significantly streamlining the design process. SPiFOL's real-time capabilities allow it to 
be used in digital twins. \\

\noindent
\textbf{Data Availability}:
All data and code used for SPiFOL are publicly available in the open repository at \href{https://github.com/Harandi-Ali/SPiFOL}{SPiFOL}.
\\ \\
\textbf{Acknowledgements}:
The authors acknowledge the financial support of Transregional Collaborative Research Center SFB/TRR 339 with project number 453596084 funded by DFG gratefully.
Sh.R. would like to thank the Deutsche Forschungsgemeinschaft
(DFG) for the funding support provided to develop the present work in the project Cluster of
Excellence Internet of Production (project: 390621612). 
H.D. and S.R. gratefully acknowledge the funding from the European Union’s
Horizon 2020 research and innovation program under the Marie Skłodowska-Curie grant agreement No 956401 (XS-Meta). \\
We would like to thank Ningyu Zhang for his assistance during the review process. 
Sh.R. would like to express his gratitude to Dr.-Ing. Reza Najian for the fruitful and insightful discussions that contributed to this work.
\\ \\ 
\textbf{Author Statement}:
A.H.: Conceptualization, Methodology, Software, Writing - Review \& Editing. 
H.D.: Software, Methodology,  Writing - Review \& Editing. 
K.L: Supervision, Review \& Editing.
S.R.: Funding, Supervision, Review \& Editing.
Sh.R.: Methodology, Writing - Review, Supervision \& Editing.
\\ 
\appendix
\section{Sample generation}
\label{sec:data}
This work presents a fully physics-based methodology. To ensure that the proposed methodology is applicable to a wide range of microstructure shapes, it is essential to diversify the dataset by incorporating a comprehensive range of samples. One of the challenges in the current study is to maintain the periodicity of the microstructure to ensure the performance of the FFT solver algorithm, which is also used in the training of SPiFOL. 

The value of the material phase, denoted by the variable $\phi$, is varied between the values $0$ and $1$.  Subsequently, two different phases are considered for the determination of the material parameters, which leads to the following formulation for the Lame constants for the material phase $\phi$ as 
\begin{equation}
\begin{aligned}
\label{eq:lame_phi}
\lambda(\phi)\,&= \lambda_f\,\phi\,+\,(1-\phi)\,\lambda_m \\
\mu(\phi)\,&= \mu_f\,\phi\,+\,(1-\phi)\,\mu_m.
\end{aligned}
\end{equation}

The latter is determined based on the weakest and strongest phases present. The maximum phase contrast ratio between the material parameters of different phases is defined as $r\,=\,\lambda_f/\lambda_m \,=\, \mu_f/\mu_m$ and can be selected to apply to different data sets. 
During training, a constant phase ratio within the microstructures should be used to ensure the applicability of the training strategy discussed in section\,\ref{sec:spifol}. 

For the 2D case, we consider both dual-phase and Fourier-based samples, whereas, for the 3D case, only the Fourier-based samples are generated. 
\subsection{2D samples}
\subsubsection{Dual-phase dataset}
\label{2D_small_samples_dualphase}
\begin{figure}[H] 
  \centering
  \includegraphics[width=1.0\linewidth]{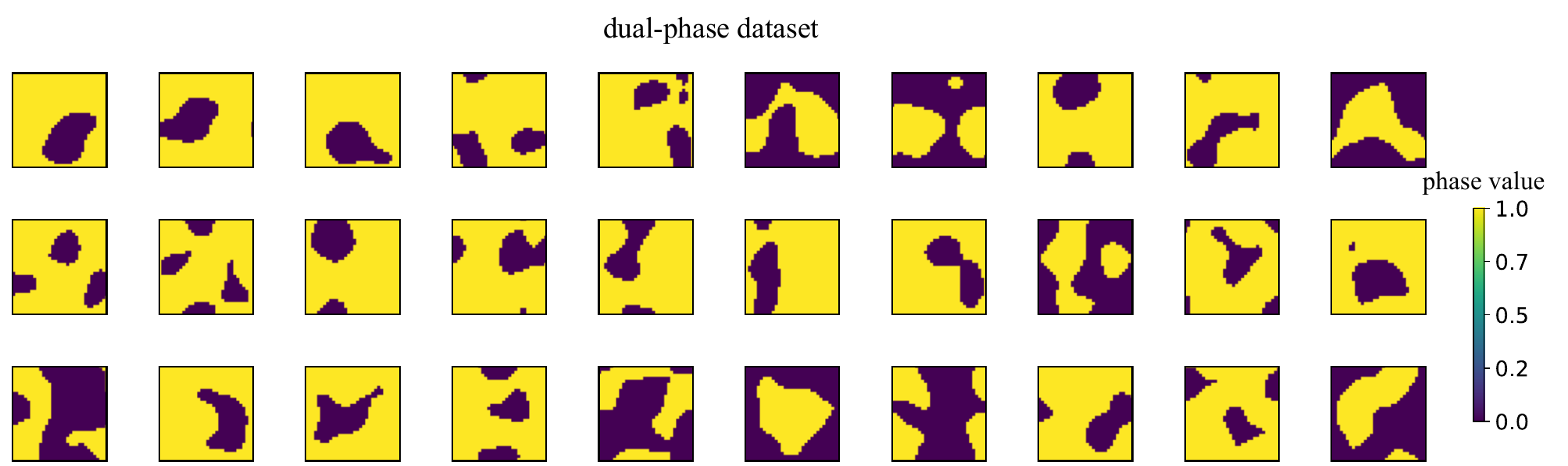}
  \caption{Randomly selected dual-phase microstructures in which phase value changes from $0$ to $1$.}
\label{fig:dual_phase}
\end{figure}

The dual-phase dataset considered in this work mimics the microstructure of steel and other high-strength alloys that fall into this category.
We have attempted to exploit the ferrite matrix with varying fractions of martensite, ranging from $0.4$ to $0.8$. To generate the corresponding data set, the grid of $32$ by $32$ equidistant points is considered. We apply a Gaussian filter with a standard deviation of $4.0$ to create clusters that can be interpreted as islands of martensite. Fig.\,\ref{fig:dual_phase} shows $30$ randomly selected cases from the dataset. Finally, the central part of the smooth extended grid is taken and checked for periodicity. If the periodicity is satisfied, the microstructure is added to the training data set. The training data set contains $8000$ samples.

\subsubsection{Fourier-based samples and reduced parametric space}
\label{Fourierbased_samples}
The Fourier-based approach combines specific frequencies with random amplitudes. To further diversify the microstructures, the \emph{sigmoid} function ($sigmoid(x)\,=\,{1}/\left(1\,+\,e^{-x}\right)$) is used, which modifies the slopes of the phase variation through various parameters. The final form of the given method is summarized as
\begin{equation}
\begin{aligned}
\label{eq:phi_sigmoid}
\bm{\phi} &= \dfrac{sigmoid(t_1(\bm{\phi}^\star-t_2)) + 0.05}{1+0.05},
\end{aligned}
\end{equation}
where $\bm{\phi}^\star$, the summation value of frequencies and random amplitudes, is defined as
\begin{equation}
\begin{aligned}
\label{eq:phi*}
\bm{\phi}^\star= \sum_{i=1}^{N_x} \sum_{j=1}^{N_y} a_i a_j \,\cos({2\pi \,fx_{i} \,\bm{x}})\, \cos({2\pi fy_{j}\,\bm{y}}).
\end{aligned}
\end{equation}
In Eq.\,\eqref{eq:phi_sigmoid}, $\bm{\phi}$ devotes to the final distribution of phases. $t_1$ and $t_2$ are the tuning parameters of the \emph{sigmoid} function which are given in Table \,\ref{tab:Data_gen}. 
$a_i$ and $a_j$ denote the normalized random amplitudes while $fx_{i}$ and $fy_{i}$ represent the Fourier frequencies in $x$ and $y$ directions and are integers. $\bm{x}$ and $\bm{y}$ show the coordinates of mesh points. 

\subsubsection{Fourier-based sample generation}
\label{sec:fourier_samples}
The list of parameters that are used to create $8000$ multiphase samples are listed in Table\,\ref{tab:Data_gen}.
\begin{table}[H]
\centering
\caption{Data generation parameters for Fourier-based approach}  
\label{tab:Data_gen}
\begin{footnotesize}
\begin{tabular}{ l l }
\hline
      parameters &  values \\
\hline
$t_1$  & $\left[-1,\,0.5,\,1,\,2,\,5,\,10,\,300  \right]$  \\
$t_2$  & $\left[0.02,\,0.1,\,0.5,\,1,\,2\right]$ \\
$N_x=N_y$ & 3\\
$\bm{f}_x$ &  $\left[0,\,1,\,2,\,3\right]$ \\
$\bm{f}_y$ &  $\left[0,\,1,\,2,\,3\right]$ \\
$\bm{a}_i$ and & normalized random amplitudes\\   
\hline
\end{tabular}
\end{footnotesize}
\end{table} 

\begin{figure}[htb] 
  \centering
  \includegraphics[width=1.0\linewidth]{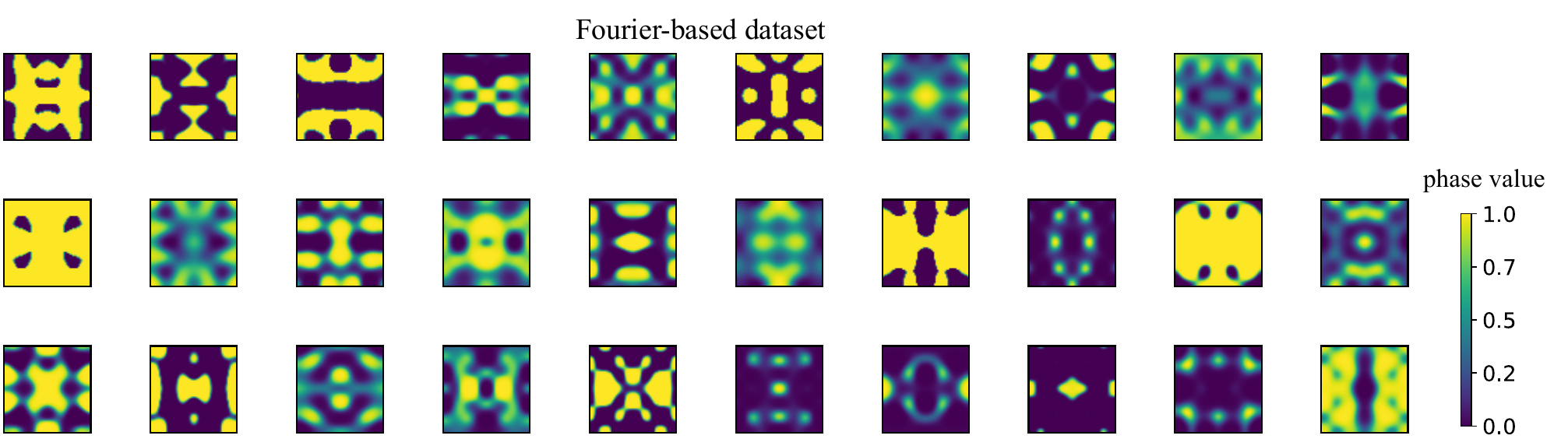}
  \caption{Randomly selected Fourier microstructures where phase changes from $0$ to $1$. They can represent multiphase materials and metamaterials.}
\label{fig:Fourier_dataset}
\end{figure}

As shown in Fig.,\ref{fig:Fourier_dataset}, the generated samples represent multiphase materials. The objective is to demonstrate the applicability of SPiFOL in predicting the mechanical behavior of arbitrary microstructures.

In the same way the 3D multiphase dataset is created by adding a third frequency, for further details please refer to \ref{appendix:3d_dataset}.

To highlight the superiority of the proposed surrogate model, SPiFOL, we also employ a data-driven version of the FNO. This data-driven model has the same architecture and parameters as the SPiFOL with FNO architecture. The training data is generated by conventional FFT solvers, allowing a direct comparison in terms of accuracy, training efficiency, and prediction performance with the SPiFOL framework. For these comparisons, the dataset is restricted to dual-phase materials.
\subsubsection{Reduced parametric space}
\label{sec:redduced_param_fourier}
The objective of this subsection is to explain how the parameters used in Fourier-based microstructure generation are leveraged to define a reduced parametric space. Utilizing our Fourier-based approach for microstructure generation, as illustrated in Fig.,\ref{fig:Fourier_dataset}, we employ 18 independent variables, detailed in Table,\ref{tab:Data_gen}. This methodology is consistent with our previous work on the concept of Finite Operator Learning (FOL) \cite{Rezaei2024fol_mech}, where it was used to prevent a significant increase in the number of network parameters.
\citet{li2023geofno} constructs a latent space for arbitrary domains using a uniform grid. However, using a latent space that exists within the physical domain does not seem feasible. Meanwhile, due to the structure of the FNO architecture and its use of a reduced set of frequencies, along with the resolution invariancy of FNOs, the latter does not necessarily lead to an increase in the number of network parameters.

Moreover, \citet{kontolati2024learning} demonstrated that the accuracy of operator performance tends to decline as the dimensionality of the parameter space increases. They also highlighted the advantages of training operators within a latent space, which can improve both efficiency and generalization.

\subsection{3D samples}
\label{appendix:3d_dataset}
The samples used for training the 3D SPiFOL model are generated in the same way as in section\,\ref{sec:fourier_samples}.
\begin{figure}[H]
  \centering
  \includegraphics[width=0.99\linewidth]{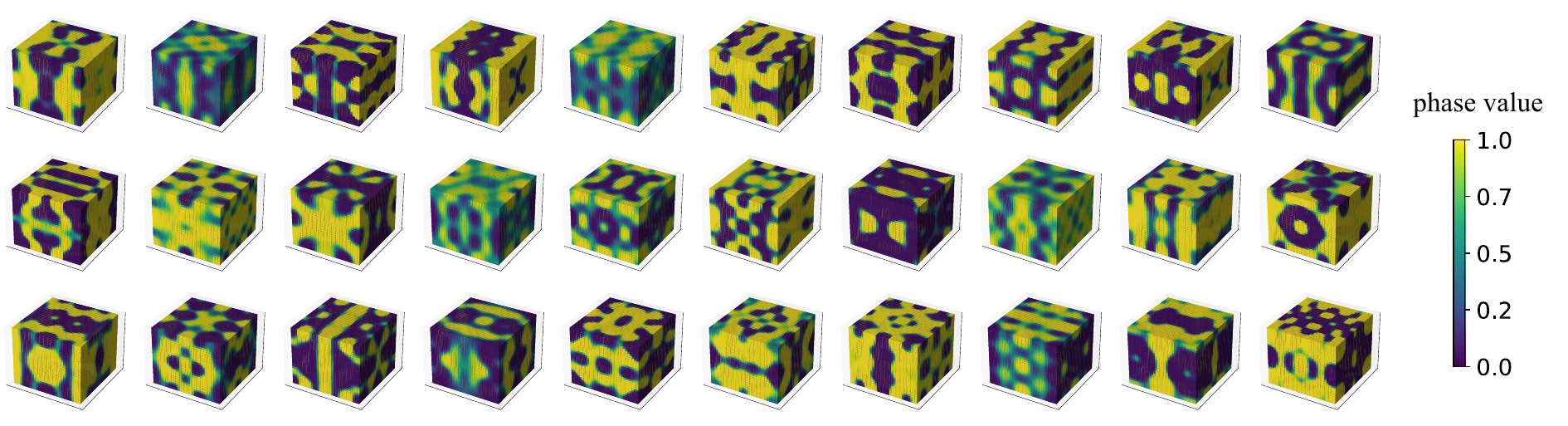}
  \caption{Randomly selected three-dimensional Fourier microstructures where phase changes from $0$ to $1$.}
\label{fig:3ddataset}
\end{figure}
The summation of frequencies for 3D is computed by adding random frequencies in $z$-direction as 
\begin{equation}
\begin{aligned}
\label{eq:phi*3d}
\bm{\phi}^\star_{3D}= \sum_{i=1}^{N_x} \sum_{j=1}^{N_y} \sum_{k=1}^{N_z} a_i a_j a_k \,\cos({2\pi \,fx_{i} \,\bm{x}})\, \cos({2\pi fy_{j}\,\bm{y}})\, \cos({2\pi fz_{k}\,\bm{z}}).
\end{aligned}
\end{equation}
In Eq.\,\eqref{eq:phi*3d}$\bm{f}_z$ three different values of $[0,\,1,\,2]$ are chosen. The final distribution of the phases is calculated by Eq.\,\eqref{eq:phi_sigmoid} and by including the remaining parameters as in Table\,\ref{tab:phase_params}.
\section{FFT-based homogenization}
\subsection{Small deformation setup}
\label{sec:FFT_algorithms}
Following the basic scheme of Mulinec and Suquet \cite{moulinec1998numerical}, the FFT-based homogenization for the small deformation setup is started by the additive split of the strain field $\bm{\varepsilon}(\bm{x})$ into the given macroscopic averaged strain field $\bar{\bm{\varepsilon}}$ and a fluctuation contribution $\widetilde{\bm{\varepsilon}}(\bm{x})$ as
\begin{equation}
\label{eq:1}
    \bm{\varepsilon} (\bm{x}) \,=\, \bar{\bm{\varepsilon}} + \widetilde{\bm{\varepsilon}} ( \bm{x} ).
\end{equation}
The constitutive relationship between stress and strain fields can be expressed by incorporating a homogeneous reference medium with stiffness $\mathbb{C}^0$, leading to the following expression for the stress field as
\begin{equation}
\label{eq:2}
    \bm{\sigma} \,=\, \mathbb{C}^0 : (\bar{\bm{\varepsilon}} \,+\, \widetilde{\bm{\varepsilon}}(\bm{x})) + \bm{\tau}(\bm{x}),
\end{equation}
where $\bm{\tau}(\bm{x})$ is the polarization stress, 
representing the deviation between the actual material stiffness $\mathbb{C}$ and the reference medium $\mathbb{C}^0$, defined as
\begin{equation}
\label{eq:3}
    \bm{\tau} (\bm{x}) = \left(\mathbb{C}(\bm{x})-\mathbb{C}^0\right):\left(\bar{\bm{\varepsilon}} + \widetilde{\bm{\varepsilon}} (\bm{x}) \right).
\end{equation}
Since the term $\mathbb{C}^0 : \bar{\bm{\varepsilon}}$ is constant, the balance of linear momentum will take the form
\begin{equation}
\label{eq:4}
    \operatorname{div} \left( \mathbb{C}^0: \widetilde{\bm{\varepsilon}} (\bm{x}) \right) + \operatorname{div} \bm{\tau}(\bm{x})=\bm{0}.
\end{equation}
Employing the Green's function approach, Eq.\,\eqref{eq:4} can be converted to the integral equation
\begin{equation}
\label{eq:5}
    \widetilde{\bm{\varepsilon}}(\bm{x}) = -\int_{\Omega} \bbGamma^0 \left(\bm{x}-\bm{x}^{\prime}\right): \bm{\tau} \left(\bm{x}^{\prime}\right) \mathrm{d} \bm{x}^{\prime},
\end{equation}
which is commonly referred to as the Lippmann-Schwinger equation. In Eq.\,\eqref{eq:5}, $\bbGamma^0$ is the Lippmann-Schwinger operator in the small strain regime and is expressed by the following form in the Fourier space:
\begin{equation}
\label{eq:LippmanSch.Oper}
    \hat{\Gamma}_{ijkl}^0 (\bm{\xi}) = \xi_l \xi_j \hat{G}^{0}_{ki}\left(\bm{\xi}\right).
\end{equation}
Here, $\hat{(\cdot)}$ is used to represent a quantity in the Fourier space, $\bm{\xi}$ is the frequency vector, and $\hat{\bm{G}}^0$ denotes the Green's function, which can be written in terms of the material stiffness $\mathbb{C}^0$ as
\begin{equation}
    \hat{G}^{0}_{ki} \left(\bm{\xi}\right) = \left[ C^{0}_{ijkl} \xi_{l} \xi{j} \right]^{-1}.
\end{equation}
By transferring it into the Fourier space, the convolution integral of Eq. \eqref{eq:LippmanSch.Oper} can be solved as
\begin{equation}
\label{eq:7}
    \hat{\widetilde{\bm{\varepsilon}}}(\bm{\xi})= -\hat{\bbGamma}^{0}(\bm{\xi}) : \hat{\bm{\tau}}(\bm{\xi}),
\end{equation}
in the Fourier space.
Finally, the total strain field in the real space is obtained by performing inverse Fourier transform $\mathcal{F}^{-1}(\cdot)$ on the fluctuation field $\hat{\widetilde{\bm{\varepsilon}}}(\bm{\xi})$ and including the contribution of the average field $\bar{\bm{\varepsilon}}$ as
\begin{equation}
\label{eq:8}
    {\bm{\varepsilon}}(\bm{x})= \bar{\bm{\varepsilon}} - \mathcal{F}^{-1}\left(\hat{\bbGamma}^{0}(\bm{\xi}) : \hat{\bm{\tau}}(\bm{\xi})\right).
\end{equation}

The choice of Fourier space shape function $\bm{\xi}$ directly impacts the Lippman-Schwinger Operator $\hat{\bbGamma}^{0}$ which is derived based on the reference medium. This operator updates the fluctuation strain fields while ensuring equilibrium in Fourier space. As a result, through iterative solution methods, the given boundary value problem under a certain macroscopic strain is solved.

\citet{schneider2021review} investigated the effects of different finite difference schemes. The second-order central difference scheme demonstrated robust performance, fast convergence, and simplicity in computing spatial gradients in Fourier space. In this scheme, the component of the Fourier shape function  $\bm{\xi}$ (frequency vector) is written as 
\begin{equation}
\label{eq:xi}
\xi_m\,=\,\mathrm{i}\sin(2\pi\,\mathrm{i}q_m/N_m)\,\dfrac{N_m}{L_m}\qquad \text{in which} \qquad q_m\,=\,-N_m/2,\,\dots,\,N_m/2-1.
\end{equation}
In Eq.\,\eqref{eq:xi},  $N_m$ is the number of grid points in each direction and $L_m$ shows the corresponding length of that dimension.

\subsection{Finite deformation setup}
The extension of the basic scheme to finite deformation is proposed by \cite{lahellec2003analysis}. By formulating equilibrium in the reference configuration by employing first Piola Kirchhoff stress tensor and by changing the definition of polarization stress tensor Eq.\,\eqref{eq:8} is reformulated to 
\begin{equation}
\label{eq:F_update}
    {\bm{F}}(\bm{X})= \bar{\bm{F}} - \mathcal{F}^{-1}\left(\hat{\bbGamma}^{0,F}(\bm{\xi}) : \hat{\bm{\tau}}^F(\bm{\xi})\right).
\end{equation}
In Eq.\,\eqref{eq:F_update}, $\hat{\bbGamma}^{0,F}$ stands for the Lippmann-Schwinger operator for finite deformation, which, unlike the small strain case (i.e., Eq. \eqref{eq:LippmanSch.Oper}), only has major symmetry and is computed by  
\begin{equation}
\label{eq:LippmanSch.OperF}
    \hat{\Gamma}_{ijkl}^{0,F} (\bm{\xi}) = \hat{G}^{0}_{ik} \left(\bm{\xi}\right) \xi_j \xi_l \Big|_{(ik)(jh)},
\end{equation}
where the notation $(ik)(jh)$ signifies that only the indices $i, k$ and $j, h$ undergo symmetrization. The polarization stress for finite deformation is also calculated as 
\begin{equation}
\label{eq:tau_F}
    \bm{\tau} ^F(\bm{X}) = \bm{P}\left(\bm{X}, \bm{F(\bm{X})}\right)\,-\,\mathbb{C}^0:F(\bm{X}).
\end{equation}
In Eqs.\,\eqref{eq:F_update}, \eqref{eq:tau_F}, $\bm{X}$ denotes to the material points in undeformed setup.  

In the SPiFOL framework, other Fourier-based shape functions can also be employed (higher order discretization scheme) without sacrificing computational efficiency during either training or evaluation. This flexibility is possible because, based on the resolution used to impose physical constraints, these shape functions only affect the Lippmann-Schwinger operator, which is constructed once before training, see \cite{schneider2021review, lucarini2021fft} for more details.

The reference material stiffness matrix $\mathbb{C}^0$ is determined by material parameters in the stiffest and softest phases available in the microstructures present in the dataset \cite{schneider2021review}. 
Moreover, Eq.\,\eqref{eq:LippmanSch.OperF} shows the Lippmann-Schwinger operator is defined by having the medium material parameters (the minimum and maximum material parameters are employed in this case). The latter results in a fixed operator for all of the samples which share the same phase contrast value. 

\section{Study over the number of Fourier modes and Fourier layers on the FNO performance} 
\label{appendix:fno_study}
In this section, we want to study the number of Fourier layers and the number of Fourier modes used in SPiFOL. To ensure the fast evaluation of SPiFOL, we aim at employing the fewest network parameters as possible.

\begin{figure}[htb] 
  \centering
  \includegraphics[width=0.95\linewidth]{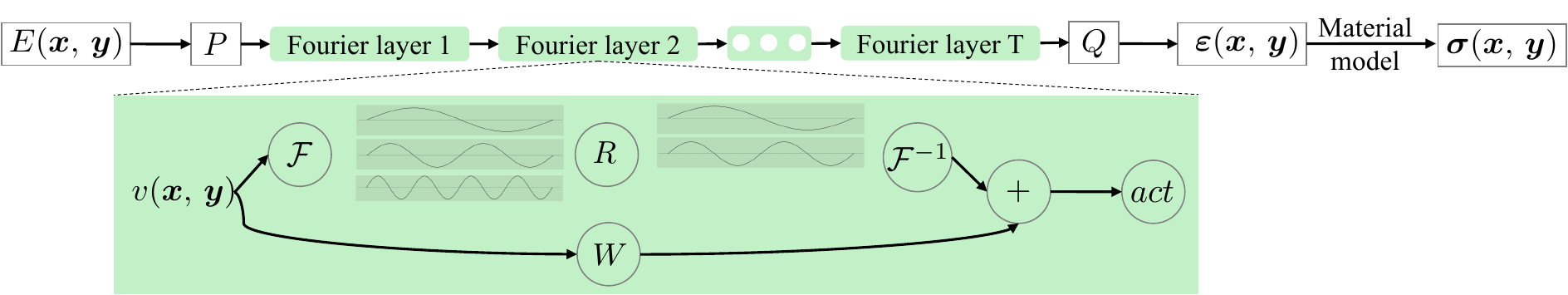}
  \caption{The structure of FNO, $P$ brings the input function to a higher dimension. Several Fourier layers are applied and then $Q$ brings back the output of the last Fourier layer to to desired output (strains in our case), adopted from \cite{li2020fourier}. Bottom: the detailed view of each Fourier layer. The higher dimension input $v(\bm{x},\,\bm{y})$ is brought to Fourier space by $\mathcal{F}$. ${R}$ truncates the higher Fourier modes and $\mathcal{F}^{-1}$ brings the output to real space. $W$ is the local linear transform of real input which is summed by the upper output and passed to the activation function.} 
  \label{fig:FNO}
\end{figure}

\begin{figure}[H] 
  \centering
  \includegraphics[width=0.87\linewidth]{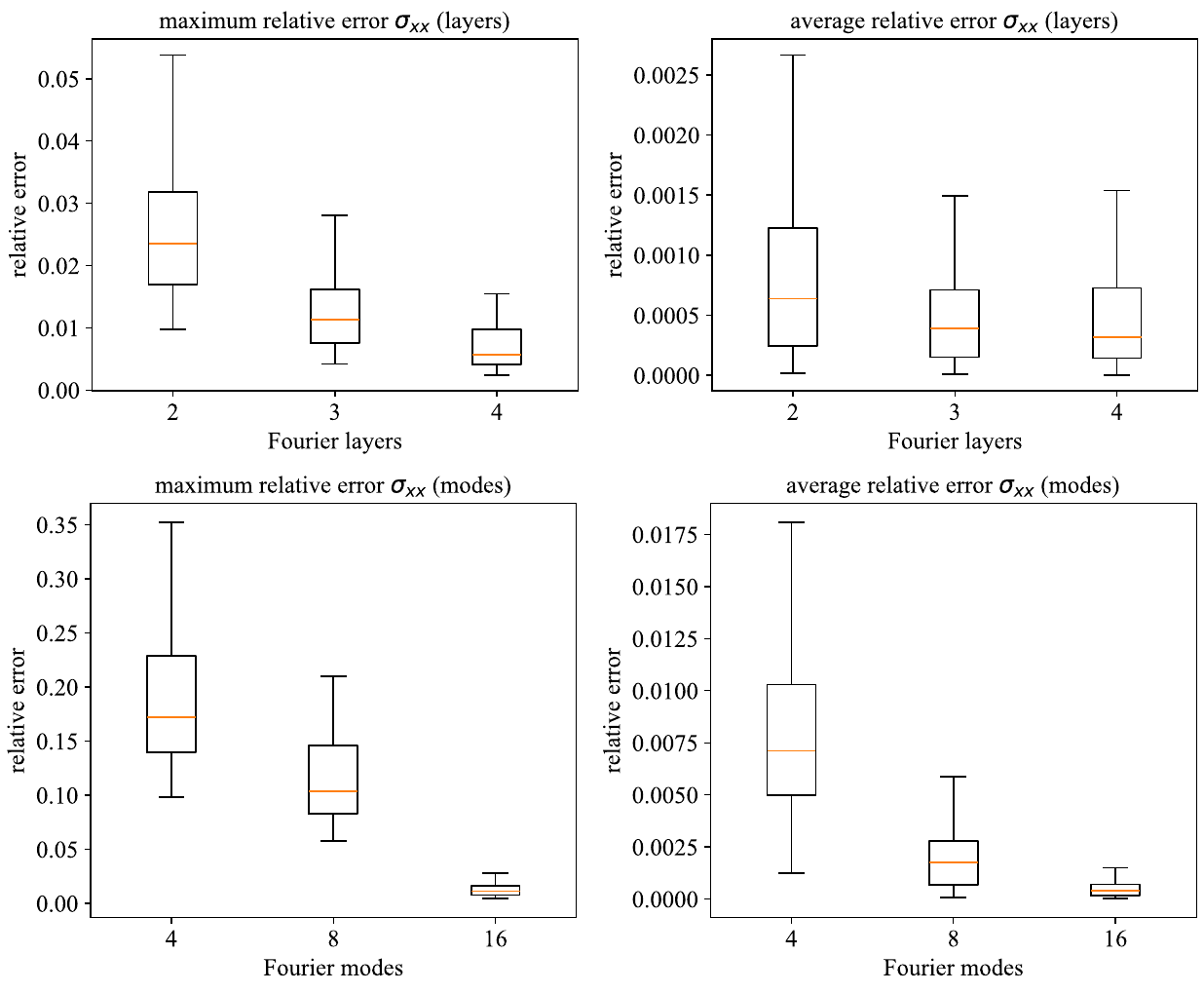}
  \caption{Architecture study of SPiFOL with FNO. On the top, the number of Fourier layers is studied and on the bottom, the number of Fourier modes is studied.}
\label{fig:arch_fno}
\end{figure}

We observe increasing the number of modes is more critical than the number of layers, see Fig.\,\ref{fig:arch_fno}. To ensure the accuracy of the network and decrease the errors $16$ modes are selected. However, we select $3$ Fourier layers to maintain the low evaluation time for the network and also prevent overfitting.
The dataset to perform this study is a dual-phase dataset and we consider the phase contrast ratio of $5$ to perform training, the training is done for $30000$ for all cases. 
\section{Single microstructure solver, on-the-fly (OTF) model} 
\label{appendix:otf}
In the case of a single microstructure, the training of SPiFOL is equivalent to solving the problem under a prescribed macroscopic strain.

\begin{figure}[H] 
  \centering
  \vspace{-5mm}
  \includegraphics[width=0.75\linewidth]{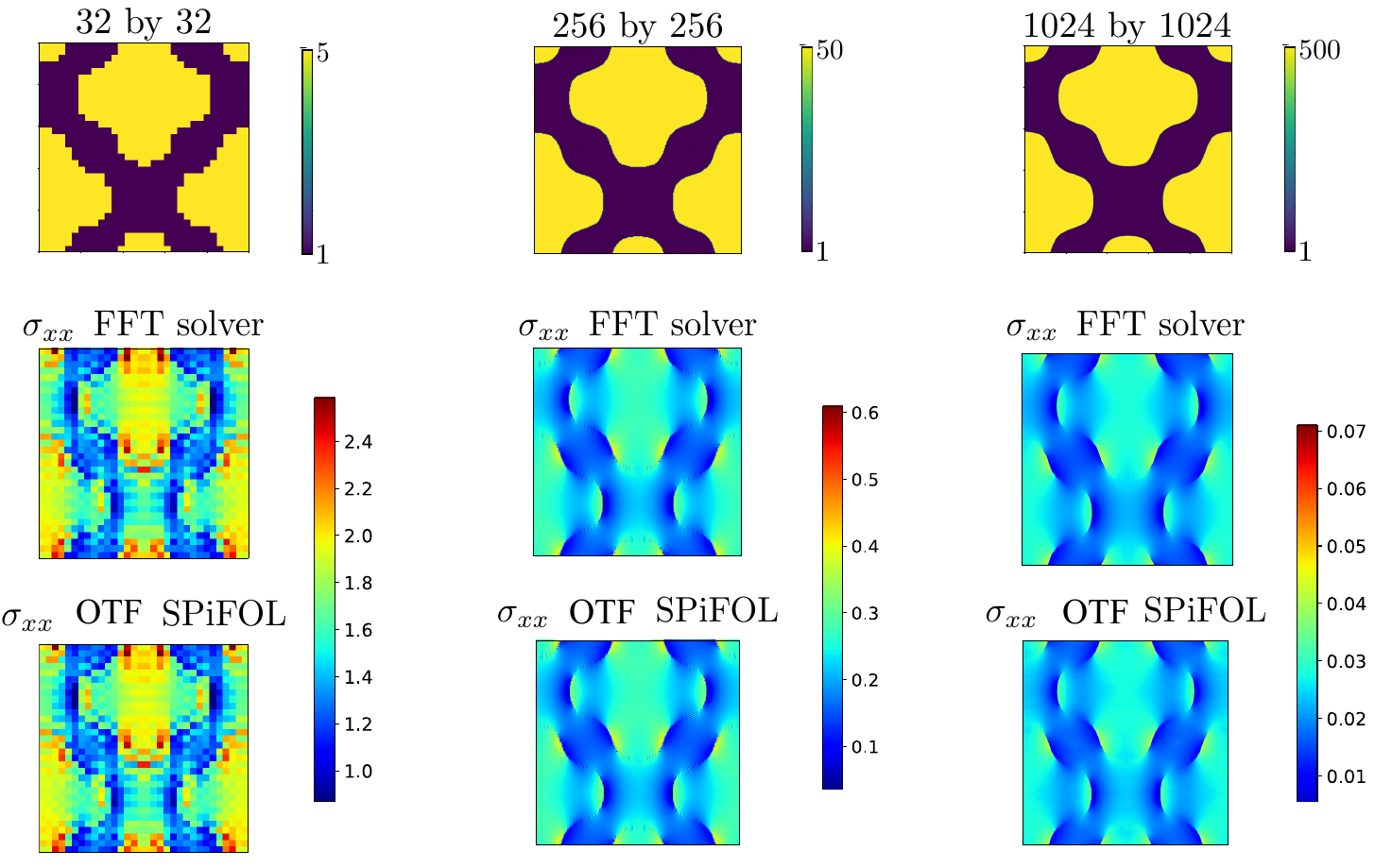}
  \caption{FFT solver solution compared to OTF SPiFOL at different resolutions and various phase contrast ratios.}
\label{fig:onf_phase}
\end{figure}

Achieving a certain level of accuracy with SPiFOL typically requires more iterations—approximately eight times more—compared to conventional FFT-based solvers. 
However, this iteration count is highly sensitive to factors such as the phase contrast ratio, network initialization, and the choice of optimizer. Fig.\,\ref{fig:onf_phase} presents a comparison between results obtained using OTF SPiFOL and a conventional FFT solver employing the Fourier-Galerkin method, under a phase contrast of 500. In Fig.~\ref{fig:onf_phase}, results are shown for increasing resolutions from left to right: $32 \times 32$, $256 \times 256$, and $1024 \times 1024$. Corresponding phase contrasts are 5, 50, and 500, respectively.

\begin{figure}[H]
  \centering
    \vspace{-5mm}
  \includegraphics[width=0.75\linewidth]{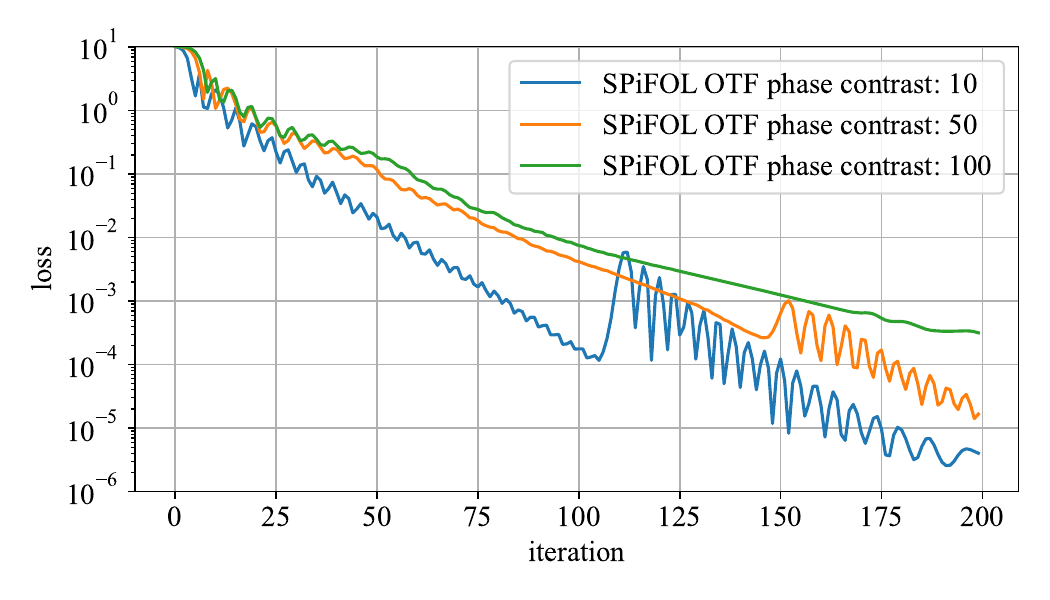}
    \vspace{-5mm}
  \caption{The loss decay of on-the-fly model (single microstructure) for three different phase contrast ratios.}.
\label{fig:otf}
\end{figure}

Fig.\,\ref{fig:otf} also shows the loss decay of OTF SPiFOL for solving a microstructure for different phase contrast ratios.
\bibliography{Ref}

\begin{thebibliography}{73}
\providecommand{\natexlab}[1]{#1}
\providecommand{\url}[1]{\texttt{#1}}
\expandafter\ifx\csname urlstyle\endcsname\relax
  \providecommand{\doi}[1]{doi: #1}\else
  \providecommand{\doi}{doi: \begingroup \urlstyle{rm}\Url}\fi

\bibitem[Liu et~al.(2022)Liu, Li, and Park]{liu2022eighty}
Wing~Kam Liu, Shaofan Li, and Harold~S Park.
\newblock Eighty years of the finite element method: Birth, evolution, and future.
\newblock \emph{Archives of Computational Methods in Engineering}, 29\penalty0 (6):\penalty0 4431--4453, 2022.

\bibitem[Sod(1978)]{sod1978survey}
Gary~A. Sod.
\newblock A survey of several finite difference methods for systems of nonlinear hyperbolic conservation laws.
\newblock \emph{Journal of computational physics}, 27\penalty0 (1):\penalty0 1--31, 1978.

\bibitem[Shen et~al.(2011)Shen, Tang, and Wang]{shen2011spectral}
Jie Shen, Tao Tang, and Li-Lian Wang.
\newblock \emph{Spectral methods: algorithms, analysis and applications}, volume~41.
\newblock Springer Science \& Business Media, 2011.

\bibitem[Karniadakis et~al.(2021)Karniadakis, Kevrekidis, Lu, Perdikaris, Wang, and Yang]{karniadakis2021physics}
George~E. Karniadakis, Ioannis~G. Kevrekidis, Lu~Lu, Paris Perdikaris, Sifan Wang, and Liu Yang.
\newblock Physics-informed machine learning.
\newblock \emph{Nature Reviews Physics}, 3\penalty0 (6):\penalty0 422--440, 2021.

\bibitem[Rao et~al.(2021)Rao, Sun, and Liu]{Rao2021}
Chengping Rao, Hao Sun, and Yang Liu.
\newblock Physics-informed deep learning for computational elastodynamics without labeled data.
\newblock \emph{Journal of Engineering Mechanics}, 147\penalty0 (8):\penalty0 04021043, 2021.

\bibitem[Rezaei et~al.(2022)Rezaei, Harandi, Moeineddin, Xu, and Reese]{REZAEI2022PINN}
Shahed Rezaei, Ali Harandi, Ahmad Moeineddin, Bai-Xiang Xu, and Stefanie Reese.
\newblock A mixed formulation for physics-informed neural networks as a potential solver for engineering problems in heterogeneous domains: Comparison with finite element method.
\newblock \emph{Computer Methods in Applied Mechanics and Engineering}, 401:\penalty0 115616, 2022.

\bibitem[Harandi et~al.(2024)Harandi, Moeineddin, Kaliske, Reese, and Rezaei]{Harandi2023}
Ali Harandi, Ahmad Moeineddin, Michael Kaliske, Stefanie Reese, and Shahed Rezaei.
\newblock Mixed formulation of physics-informed neural networks for thermo-mechanically coupled systems and heterogeneous domains.
\newblock \emph{International Journal for Numerical Methods in Engineering}, 125\penalty0 (4):\penalty0 e7388, 2024.

\bibitem[Lu et~al.(2021{\natexlab{a}})Lu, Pestourie, Yao, Wang, Verdugo, and Johnson]{Luinverse}
Lu~Lu, Rapha\"{e}l Pestourie, Wenjie Yao, Zhicheng Wang, Francesc Verdugo, and Steven~G. Johnson.
\newblock Physics-informed neural networks with hard constraints for inverse design.
\newblock \emph{SIAM Journal on Scientific Computing}, 43\penalty0 (6):\penalty0 B1105--B1132, 2021{\natexlab{a}}.

\bibitem[Chen et~al.(2020)Chen, Lu, Karniadakis, and Negro]{Chen2020}
Yuyao Chen, Lu~Lu, George~E. Karniadakis, and Luca~Dal Negro.
\newblock Physics-informed neural networks for inverse problems in nano-optics and metamaterials.
\newblock \emph{Opt. Express}, 28\penalty0 (8):\penalty0 11618--11633, Apr 2020.

\bibitem[Wang et~al.(2022{\natexlab{a}})Wang, Yu, and Perdikaris]{WANG2022why}
Sifan Wang, Xinling Yu, and Paris Perdikaris.
\newblock When and why {PINN}s fail to train: A neural tangent kernel perspective.
\newblock \emph{Journal of Computational Physics}, 449:\penalty0 110768, 2022{\natexlab{a}}.

\bibitem[Krishnapriyan et~al.(2021)Krishnapriyan, Gholami, Zhe, Kirby, and Mahoney]{krishnapriyan2021characterizing}
Aditi Krishnapriyan, Amir Gholami, Shandian Zhe, Robert Kirby, and Michael~W Mahoney.
\newblock Characterizing possible failure modes in physics-informed neural networks.
\newblock \emph{Advances in neural information processing systems}, 34:\penalty0 26548--26560, 2021.

\bibitem[Lu et~al.(2021{\natexlab{b}})Lu, Jin, Pang, Zhang, and Karniadakis]{Lu2021}
Lu~Lu, Pengzhan Jin, Guofei Pang, Zhongqiang Zhang, and George~E. Karniadakis.
\newblock Learning nonlinear operators via {DeepONet} based on the universal approximation theorem of operators.
\newblock \emph{Nature Machine Intelligence}, 2021{\natexlab{b}}.

\bibitem[Koric et~al.(2024)Koric, Viswantah, Abueidda, Sobh, and Khan]{koric2024deep}
Seid Koric, Asha Viswantah, Diab~W. Abueidda, Nahil~A. Sobh, and Kamran Khan.
\newblock Deep learning operator network for plastic deformation with variable loads and material properties.
\newblock \emph{Engineering with Computers}, 40\penalty0 (2):\penalty0 917--929, 2024.

\bibitem[He et~al.(2023)He, Koric, Kushwaha, Park, Abueidda, and Jasiuk]{he2023novel}
Junyan He, Seid Koric, Shashank Kushwaha, Jaewan Park, Diab~W. Abueidda, and Iwona Jasiuk.
\newblock Novel {DeepONet} architecture to predict stresses in elastoplastic structures with variable complex geometries and loads.
\newblock \emph{Computer Methods in Applied Mechanics and Engineering}, 415:\penalty0 116277, 2023.

\bibitem[Goswami et~al.(2022)Goswami, Yin, Yu, and Karniadakis]{GOSWAMI2022114587}
Somdatta Goswami, Minglang Yin, Yue Yu, and George~E. Karniadakis.
\newblock A physics-informed variational {DeepONet} for predicting crack path in quasi-brittle materials.
\newblock \emph{Computer Methods in Applied Mechanics and Engineering}, 391:\penalty0 114587, 2022.
\newblock ISSN 0045-7825.

\bibitem[Wang et~al.(2022{\natexlab{b}})Wang, Wang, and Perdikaris]{wang2022improved}
Sifan Wang, Hanwen Wang, and Paris Perdikaris.
\newblock Improved architectures and training algorithms for deep operator networks.
\newblock \emph{Journal of Scientific Computing}, 92\penalty0 (2):\penalty0 35, 2022{\natexlab{b}}.

\bibitem[Haghighat et~al.(2024)Haghighat, bin Waheed, and Karniadakis]{haghighat2024deeponet}
Ehsan Haghighat, Umair bin Waheed, and George~E. Karniadakis.
\newblock En-{DeepONet}: An enrichment approach for enhancing the expressivity of neural operators with applications to seismology.
\newblock \emph{Computer Methods in Applied Mechanics and Engineering}, 420:\penalty0 116681, 2024.

\bibitem[Lu et~al.(2022)Lu, Meng, Cai, Mao, Goswami, Zhang, and Karniadakis]{LU2022114778}
Lu~Lu, Xuhui Meng, Shengze Cai, Zhiping Mao, Somdatta Goswami, Zhongqiang Zhang, and George~E. Karniadakis.
\newblock A comprehensive and fair comparison of two neural operators (with practical extensions) based on fair data.
\newblock \emph{Computer Methods in Applied Mechanics and Engineering}, 393:\penalty0 114778, 2022.

\bibitem[Kontolati et~al.(2024)Kontolati, Goswami, E.~Karniadakis, and Shields]{kontolati2024learning}
Katiana Kontolati, Somdatta Goswami, George E.~Karniadakis, and Michael~D Shields.
\newblock Learning nonlinear operators in latent spaces for real-time predictions of complex dynamics in physical systems.
\newblock \emph{Nature Communications}, 15\penalty0 (1):\penalty0 5101, 2024.

\bibitem[He et~al.(2024)He, Koric, Abueidda, Najafi, and Jasiuk]{he2024geom}
Junyan He, Seid Koric, Diab~W. Abueidda, Ali Najafi, and Iwona Jasiuk.
\newblock Geom-{DeepONet}: A point-cloud-based deep operator network for field predictions on 3d parameterized geometries.
\newblock \emph{Computer Methods in Applied Mechanics and Engineering}, 429:\penalty0 117130, 2024.

\bibitem[Bahmani et~al.(2024)Bahmani, Goswami, Kevrekidis, and Shields]{bahmani2024resolution}
Bahador Bahmani, Somdatta Goswami, Ioannis~G. Kevrekidis, and Michael~D. Shields.
\newblock A resolution independent neural operator.
\newblock \emph{arXiv preprint arXiv:2407.13010}, 2024.

\bibitem[Li et~al.(2020{\natexlab{a}})Li, Kovachki, Azizzadenesheli, Liu, Bhattacharya, Stuart, and Anandkumar]{li2020neural}
Zongyi Li, Nikola~B. Kovachki, Kamyar Azizzadenesheli, Burigede Liu, Kaushik Bhattacharya, Andrew~M. Stuart, and Anima Anandkumar.
\newblock Neural operator: Graph kernel network for partial differential equations.
\newblock \emph{arXiv:2003.03485}, 2020{\natexlab{a}}.

\bibitem[Kovachki et~al.(2023)Kovachki, Li, Liu, Azizzadenesheli, Bhattacharya, Stuart, and Anandkumar]{kovachki2023neural}
Nikola~B. Kovachki, Zongyi Li, Burigede Liu, Kamyar Azizzadenesheli, Kaushik Bhattacharya, Andrew Stuart, and Anima Anandkumar.
\newblock Neural operator: Learning maps between function spaces with applications to {PDE}s.
\newblock \emph{Journal of Machine Learning Research}, 24\penalty0 (89):\penalty0 1--97, 2023.

\bibitem[Li et~al.(2023{\natexlab{a}})Li, Huang, Liu, and Anandkumar]{li2023geofno}
Zongyi Li, Daniel~Zhengyu Huang, Burigede Liu, and Anima Anandkumar.
\newblock {Fourier} neural operator with learned deformations for {PDE}s on general geometries.
\newblock \emph{Journal of Machine Learning Research}, 24\penalty0 (388):\penalty0 1--26, 2023{\natexlab{a}}.

\bibitem[Mehran et~al.(2022)Mehran, Pittie, Chakraborty, and Krishnan]{Mehran2022}
Meer Mehran, Tanu Pittie, Souvik Chakraborty, and N.~M.~Anoop Krishnan.
\newblock Learning the stress-strain fields in digital composites using {Fourier} neural operator.
\newblock \emph{iScience}, 25:\penalty0 105452, 11 2022.

\bibitem[You et~al.(2022)You, Zhang, Ross, Lee, and Yu]{YOU2022115296}
Huaiqian You, Quinn Zhang, Colton~J. Ross, Chung-Hao Lee, and Yue Yu.
\newblock Learning deep implicit {Fourier} neural operators ({IFNO}s) with applications to heterogeneous material modeling.
\newblock \emph{Computer Methods in Applied Mechanics and Engineering}, 398:\penalty0 115296, 2022.
\newblock ISSN 0045-7825.

\bibitem[Wang et~al.(2024{\natexlab{a}})Wang, Li, Yan, Du, Bai, Liu, Rabczuk, and Liu]{wang2024homogenius}
Yizheng Wang, Xiang Li, Ziming Yan, Yuqing Du, Jinshuai Bai, Bokai Liu, Timon Rabczuk, and Yinghua Liu.
\newblock {HomoGenius}: a foundation model of homogenization for rapid prediction of effective mechanical properties using neural operators.
\newblock \emph{arXiv preprint arXiv:2404.07943}, 2024{\natexlab{a}}.

\bibitem[Kovachki et~al.(2024)Kovachki, Lanthaler, and Stuart]{kovachki2024operator}
Nikola~B. Kovachki, Samuel Lanthaler, and Andrew~M. Stuart.
\newblock Operator learning: Algorithms and analysis.
\newblock \emph{arXiv preprint arXiv:2402.15715}, 2024.

\bibitem[Boullé and Townsend(2024)]{BOULLE202483}
Nicolas Boullé and Alex Townsend.
\newblock Chapter 3 - a mathematical guide to operator learning.
\newblock In Siddhartha Mishra and Alex Townsend, editors, \emph{Numerical Analysis Meets Machine Learning}, volume~25 of \emph{Handbook of Numerical Analysis}, pages 83--125. Elsevier, 2024.

\bibitem[Faroughi et~al.(2024)Faroughi, Pawar, Fernandes, Raissi, Das, Kalantari, and Mahjour]{faroughi2022physics}
Salah~A. Faroughi, Nikhil~M. Pawar, Celio Fernandes, Maziar Raissi, Subasish Das, Nima~K. Kalantari, and Seyed~Kourosh Mahjour.
\newblock Physics-guided, physics-informed, and physics-encoded neural networks and operators in scientific computing: Fluid and solid mechanics.
\newblock \emph{Journal of Computing and Information Science in Engineering}, 24\penalty0 (4):\penalty0 040802, 2024.

\bibitem[Wang et~al.(2021)Wang, Wang, and Perdikaris]{wang2021learningphysdeepon}
Sifan Wang, Hanwen Wang, and Paris Perdikaris.
\newblock Learning the solution operator of parametric partial differential equations with physics-informed {DeepONet}s.
\newblock \emph{Science advances}, 7\penalty0 (40):\penalty0 eabi8605, 2021.

\bibitem[Li et~al.(2023{\natexlab{b}})Li, Bazant, and Zhu]{li2023phasedeeponet}
Wei Li, Martin~Z. Bazant, and Juner Zhu.
\newblock {Phase-Field DeepONet}: Physics-informed deep operator neural network for fast simulations of pattern formation governed by gradient flows of free-energy functionals.
\newblock \emph{Computer Methods in Applied Mechanics and Engineering}, 416:\penalty0 116299, 2023{\natexlab{b}}.

\bibitem[Li et~al.(2023{\natexlab{c}})Li, Zheng, Kovachki, Jin, Chen, Liu, Azizzadenesheli, and Anandkumar]{li2023physicsinformed}
Zongyi Li, Hongkai Zheng, Nikola~B. Kovachki, David Jin, Haoxuan Chen, Burigede Liu, Kamyar Azizzadenesheli, and Anima Anandkumar.
\newblock Physics-informed neural operator for learning partial differential equations.
\newblock \emph{arXiv:2111.03794}, 2023{\natexlab{c}}.

\bibitem[Khorrami et~al.(2024)Khorrami, Goyal, Mianroodi, Svendsen, Benner, and Raabe]{khorrami2024divergence}
Mohammad~S. Khorrami, Pawan Goyal, Jaber~R. Mianroodi, Bob Svendsen, Peter Benner, and Dierk Raabe.
\newblock Divergence-free neural operators for stress field modeling in polycrystalline materials.
\newblock \emph{arXiv preprint arXiv:2408.15408}, 2024.

\bibitem[Kapoor et~al.(2022)Kapoor, Mianroodi, Khorrami, Siboni, and Svendsen]{kapoor2022comparison}
Sarthak Kapoor, Jaber~R. Mianroodi, Mohammad~S. Khorrami, Nima~S. Siboni, and Bob Svendsen.
\newblock Comparison of two artificial neural networks trained for the surrogate modeling of stress in materially heterogeneous elastoplastic solids.
\newblock \emph{arXiv preprint arXiv:2210.16994}, 2022.

\bibitem[Rashid et~al.(2023)Rashid, Chakraborty, and Krishnan]{RASHID2023105444}
Meer~Mehran Rashid, Souvik Chakraborty, and N.M.~Anoop Krishnan.
\newblock Revealing the predictive power of neural operators for strain evolution in digital composites.
\newblock \emph{Journal of the Mechanics and Physics of Solids}, 181:\penalty0 105444, 2023.

\bibitem[Goswami et~al.(2020)Goswami, Anitescu, Chakraborty, and Rabczuk]{GOSWAMI20TPF}
Somdatta Goswami, Cosmin Anitescu, Souvik Chakraborty, and Timon Rabczuk.
\newblock Transfer learning enhanced physics informed neural network for phase-field modeling of fracture.
\newblock \emph{Theoretical and Applied Fracture Mechanics}, 106:\penalty0 102447, 2020.

\bibitem[Rosofsky et~al.(2023)Rosofsky, Al~Majed, and Huerta]{rosofsky2023applications}
Shawn~G. Rosofsky, Hani Al~Majed, and E.~A. Huerta.
\newblock Applications of physics informed neural operators.
\newblock \emph{Machine Learning: Science and Technology}, 4\penalty0 (2):\penalty0 025022, 2023.

\bibitem[Chiu et~al.(2022)Chiu, Wong, Ooi, Dao, and Ong]{CHIU2022114909}
Pao-Hsiung Chiu, Jian~Cheng Wong, Chinchun Ooi, My~Ha Dao, and Yew-Soon Ong.
\newblock Can-pinn: A fast physics-informed neural network based on coupled-automatic–numerical differentiation method.
\newblock \emph{Computer Methods in Applied Mechanics and Engineering}, 395:\penalty0 114909, 2022.
\newblock ISSN 0045-7825.

\bibitem[Ren et~al.(2022)Ren, Rao, Liu, Wang, and Sun]{REN2022114399}
Pu~Ren, Chengping Rao, Yang Liu, Jian-Xun Wang, and Hao Sun.
\newblock Phycrnet: Physics-informed convolutional-recurrent network for solving spatiotemporal {PDE}s.
\newblock \emph{Computer Methods in Applied Mechanics and Engineering}, 389:\penalty0 114399, 2022.

\bibitem[Zhao et~al.(2023)Zhao, Gong, Zhang, Yao, and Chen]{ZHAO2023105516}
Xiaoyu Zhao, Zhiqiang Gong, Yunyang Zhang, Wen Yao, and Xiaoqian Chen.
\newblock Physics-informed convolutional neural networks for temperature field prediction of heat source layout without labeled data.
\newblock \emph{Engineering Applications of Artificial Intelligence}, 117:\penalty0 105516, 2023.

\bibitem[Zhang et~al.(2023)Zhang, Yan, Liu, Zhang, Han, and Wang]{ZHANG2023111919}
Zhao Zhang, Xia Yan, Piyang Liu, Kai Zhang, Renmin Han, and Sheng Wang.
\newblock A physics-informed convolutional neural network for the simulation and prediction of two-phase darcy flows in heterogeneous porous media.
\newblock \emph{Journal of Computational Physics}, 477:\penalty0 111919, 2023.

\bibitem[Phillips et~al.(2023)Phillips, Heaney, Chen, Buchan, and Pain]{Phillips2023}
Toby~R.F. Phillips, Claire~E. Heaney, Boyang Chen, Andrew~G. Buchan, and Christopher~C. Pain.
\newblock Solving the discretised neutron diffusion equations using neural networks.
\newblock \emph{International Journal for Numerical Methods in Engineering}, 124\penalty0 (21):\penalty0 4659--4686, 2023.

\bibitem[Rezaei et~al.(2025)Rezaei, Asl, Faroughi, Asgharzadeh, Harandi, Koopas, Laschet, Reese, and Apel]{Rezaei2024fol_mech}
Shahed Rezaei, Reza~Najian Asl, Shirko Faroughi, Mahdi Asgharzadeh, Ali Harandi, Rasoul~Najafi Koopas, Gottfried Laschet, Stefanie Reese, and Markus Apel.
\newblock A finite operator learning technique for mapping the elastic properties of microstructures to their mechanical deformations, 2025.

\bibitem[Yamazaki et~al.(2024)Yamazaki, Harandi, Muramatsu, Viardin, Apel, Brepols, Reese, and Rezaei]{yamazaki2024}
Yusuke Yamazaki, Ali Harandi, Mayu Muramatsu, Alexandre Viardin, Markus Apel, Tim Brepols, Stefanie Reese, and Shahed Rezaei.
\newblock A finite element-based physics-informed operator learning framework for spatiotemporal partial differential equations on arbitrary domains.
\newblock \emph{Engineering with Computers}, pages 1--29, 2024.

\bibitem[Rezaei et~al.(2024)Rezaei, Asl, Taghikhani, Moeineddin, Kaliske, and Apel]{Rezaei2024finite}
Shahed Rezaei, Reza~Najian Asl, Kianoosh Taghikhani, Ahmad Moeineddin, Michael Kaliske, and Markus Apel.
\newblock Finite operator learning: Bridging neural operators and numerical methods for efficient parametric solution and optimization of {PDE}s.
\newblock \emph{arXiv preprint arXiv:2407.04157}, 2024.

\bibitem[Raju et~al.(2021)Raju, Tay, and Tan]{raju2021review}
Karthikayen Raju, Tong-Earn Tay, and Vincent Beng~Chye Tan.
\newblock A review of the fe 2 method for composites.
\newblock \emph{Multiscale and Multidisciplinary Modeling, Experiments and Design}, 4:\penalty0 1--24, 2021.

\bibitem[Spahn et~al.(2014)Spahn, Andrä, Kabel, and Müller]{SPAHN2014871}
Johannes Spahn, Heiko Andrä, Matthias Kabel, and Ralf Müller.
\newblock A multiscale approach for modeling progressive damage of composite materials using fast {Fourier} transforms.
\newblock \emph{Computer Methods in Applied Mechanics and Engineering}, 268:\penalty0 871--883, 2014.
\newblock ISSN 0045-7825.

\bibitem[Lucarini et~al.(2021)Lucarini, Upadhyay, and Segurado]{lucarini2021fft}
Sergio Lucarini, Manas~V. Upadhyay, and Javier Segurado.
\newblock {FFT} based approaches in micromechanics: fundamentals, methods and applications.
\newblock \emph{Modelling and Simulation in Materials Science and Engineering}, 30\penalty0 (2):\penalty0 023002, 2021.

\bibitem[Schneider(2021)]{schneider2021review}
Matti Schneider.
\newblock A review of nonlinear {FFT}-based computational homogenization methods.
\newblock \emph{Acta Mechanica}, 232\penalty0 (6):\penalty0 2051--2100, 2021.

\bibitem[Risthaus and Schneider(2024)]{Risthaus24}
Lennart Risthaus and Matti Schneider.
\newblock {FFT}-based computational micromechanics with dirichlet boundary conditions on the rotated staggered grid.
\newblock \emph{International Journal for Numerical Methods in Engineering}, 125\penalty0 (21):\penalty0 e7569, 2024.

\bibitem[Lucarini et~al.(2023)Lucarini, Dunne, and Mart{\'\i}nez-Pa{\~n}eda]{lucarini2023fft}
Sergio Lucarini, Fionn~P.E. Dunne, and Emilio Mart{\'\i}nez-Pa{\~n}eda.
\newblock An {FFT}-based crystal plasticity phase-field model for micromechanical fatigue cracking based on the stored energy density.
\newblock \emph{International Journal of Fatigue}, 172, 2023.

\bibitem[Danesh et~al.(2024)Danesh, Di~Lorenzo, Chinesta, Reese, and Brepols]{danesh2024fft}
Hooman Danesh, Daniele Di~Lorenzo, Francisco Chinesta, Stefanie Reese, and Tim Brepols.
\newblock {FFT}-based surrogate modeling of auxetic metamaterials with real-time prediction of effective elastic properties and swift inverse design.
\newblock \emph{Materials \& Design}, 248:\penalty0 113491, 2024.

\bibitem[Kumar et~al.(2020)Kumar, Vidyasagar, and Kochmann]{kumar2020assessment}
Siddhant Kumar, Ananthan Vidyasagar, and Dennis~M Kochmann.
\newblock An assessment of numerical techniques to find energy-minimizing microstructures associated with nonconvex potentials.
\newblock \emph{International Journal for Numerical Methods in Engineering}, 121\penalty0 (7):\penalty0 1595--1628, 2020.

\bibitem[Schneider et~al.(2019)Schneider, Wicht, and B{\"o}hlke]{schneider2019polarization}
Matti Schneider, Daniel Wicht, and Thomas B{\"o}hlke.
\newblock On polarization-based schemes for the {FFT}-based computational homogenization of inelastic materials.
\newblock \emph{Computational Mechanics}, 64:\penalty0 1073--1095, 2019.

\bibitem[Moulinec and Suquet(1998)]{moulinec1998numerical}
Herv{\'e} Moulinec and Pierre Suquet.
\newblock A numerical method for computing the overall response of nonlinear composites with complex microstructure.
\newblock \emph{Computer methods in applied mechanics and engineering}, 157\penalty0 (1-2):\penalty0 69--94, 1998.

\bibitem[Li et~al.(2020{\natexlab{b}})Li, Kovachki, Azizzadenesheli, Liu, Bhattacharya, Stuart, and Anandkumar]{li2020fourier}
Zongyi Li, Nikola~B. Kovachki, Kamyar Azizzadenesheli, Burigede Liu, Kaushik Bhattacharya, Andrew Stuart, and Anima Anandkumar.
\newblock {Fourier} neural operator for parametric partial differential equations.
\newblock \emph{arXiv preprint arXiv:2010.08895}, 2020{\natexlab{b}}.

\bibitem[Jacot et~al.(2018)Jacot, Gabriel, and Hongler]{jacot2018neural}
Arthur Jacot, Franck Gabriel, and Cl{\'e}ment Hongler.
\newblock Neural tangent kernel: Convergence and generalization in neural networks.
\newblock \emph{Advances in neural information processing systems}, 31, 2018.

\bibitem[Bradbury et~al.(2018)Bradbury, Frostig, Hawkins, Johnson, Leary, Maclaurin, Necula, Paszke, Vander{P}las, Wanderman-{M}ilne, and Zhang]{jax2018}
James Bradbury, Roy Frostig, Peter Hawkins, Matthew~J. Johnson, Chris Leary, Dougal Maclaurin, George Necula, Adam Paszke, Jake Vander{P}las, Skye Wanderman-{M}ilne, and Qiao Zhang.
\newblock {JAX}: composable transformations of {P}ython+{N}um{P}y programs, 2018.

\bibitem[Akiba et~al.(2019)Akiba, Sano, Yanase, Ohta, and Koyama]{optuna_2019}
Takuya Akiba, Shotaro Sano, Toshihiko Yanase, Takeru Ohta, and Masanori Koyama.
\newblock Optuna: A next-generation hyperparameter optimization framework.
\newblock In \emph{Proceedings of the 25th {ACM} {SIGKDD} International Conference on Knowledge Discovery and Data Mining}, 2019.

\bibitem[Bassey et~al.(2021)Bassey, Qian, and Li]{bassey2021survey}
Joshua Bassey, Lijun Qian, and Xianfang Li.
\newblock A survey of complex-valued neural networks.
\newblock \emph{arXiv preprint arXiv:2101.12249}, 2021.

\bibitem[Li et~al.(2024)Li, Zheng, Kovachki, Jin, Chen, Liu, Azizzadenesheli, and Anandkumar]{li2024physics}
Zongyi Li, Hongkai Zheng, Nikola~B. Kovachki, David Jin, Haoxuan Chen, Burigede Liu, Kamyar Azizzadenesheli, and Anima Anandkumar.
\newblock Physics-informed neural operator for learning partial differential equations.
\newblock \emph{ACM/JMS Journal of Data Science}, 1\penalty0 (3):\penalty0 1--27, 2024.

\bibitem[Danesh et~al.(2023)Danesh, Brepols, and Reese]{danesh2023challenges}
Hooman Danesh, Tim Brepols, and Stefanie Reese.
\newblock Challenges in two-scale computational homogenization of mechanical metamaterials.
\newblock \emph{PAMM}, 23\penalty0 (1):\penalty0 e202200139, 2023.

\bibitem[Sinha et~al.(2024)Sinha, Benton, and Emami]{sinha2024effectiveness}
Saumya Sinha, Brandon Benton, and Patrick Emami.
\newblock On the effectiveness of neural operators at zero-shot weather downscaling.
\newblock \emph{arXiv preprint arXiv:2409.13955}, 2024.

\bibitem[{de Geus} et~al.(2017){de Geus}, Vondřejc, Zeman, Peerlings, and Geers]{DEGEUS2017412}
T.W.J. {de Geus}, J.~Vondřejc, J.~Zeman, R.H.J. Peerlings, and M.G.D. Geers.
\newblock Finite strain {FFT}-based non-linear solvers made simple.
\newblock \emph{Computer Methods in Applied Mechanics and Engineering}, 318:\penalty0 412--430, 2017.
\newblock ISSN 0045-7825.

\bibitem[Kabel et~al.(2014)Kabel, B{\"o}hlke, and Schneider]{kabel2014efficient}
Matthias Kabel, Thomas B{\"o}hlke, and Matti Schneider.
\newblock Efficient fixed point and newton--krylov solvers for {FFT}-based homogenization of elasticity at large deformations.
\newblock \emph{Computational Mechanics}, 54\penalty0 (6):\penalty0 1497--1514, 2014.

\bibitem[Asl et~al.(2025)Asl, Yamazaki, Taghikhani, Muramatsu, Apel, and Rezaei]{asl2025}
Reza~Najian Asl, Yusuke Yamazaki, Kianoosh Taghikhani, Mayu Muramatsu, Markus Apel, and Shahed Rezaei.
\newblock A {Physics-Informed Meta-Learning Framework} for the {Continuous Solution} of {Parametric} {PDE}s on {Arbitrary Geometries}.
\newblock \emph{arXiv preprint arXiv:2504.02459}, 2025.

\bibitem[Xiao et~al.(2023)Xiao, Hao, Lin, Deng, and Su]{xiao2023improved}
Zipeng Xiao, Zhongkai Hao, Bokai Lin, Zhijie Deng, and Hang Su.
\newblock Improved operator learning by orthogonal attention.
\newblock \emph{arXiv preprint arXiv:2310.12487}, 2023.

\bibitem[Calvello et~al.(2024)Calvello, Kovachki, Levine, and Stuart]{calvello2024continuum}
Edoardo Calvello, Nikola~B. Kovachki, Matthew~E. Levine, and Andrew~M Stuart.
\newblock Continuum attention for neural operators.
\newblock \emph{arXiv preprint arXiv:2406.06486}, 2024.

\bibitem[Tran et~al.(2021)Tran, Mathews, Xie, and Ong]{tran2021factorized}
Alasdair Tran, Alexander Mathews, Lexing Xie, and Cheng~Soon Ong.
\newblock Factorized {Fourier} neural operators.
\newblock \emph{arXiv preprint arXiv:2111.13802}, 2021.

\bibitem[Wang et~al.(2024{\natexlab{b}})Wang, Wu, Zhang, Fang, Liang, Wu, Zimmermann, and Wang]{wang10428110}
Kun Wang, Hao Wu, Guibin Zhang, Junfeng Fang, Yuxuan Liang, Yuankai Wu, Roger Zimmermann, and Yang Wang.
\newblock { Modeling Spatio-Temporal Dynamical Systems With Neural Discrete Learning and Levels-of-Experts }.
\newblock \emph{IEEE Transactions on Knowledge \& Data Engineering}, 36\penalty0 (08):\penalty0 4050--4062, 2024{\natexlab{b}}.
\newblock ISSN 1558-2191.

\bibitem[Jiang et~al.(2025)Jiang, Li, Wang, Yang, and Wang]{jiang2025implicit}
Yuchi Jiang, Zhijie Li, Yunpeng Wang, Huiyu Yang, and Jianchun Wang.
\newblock An implicit adaptive {Fourier} neural operator for long-term predictions of three-dimensional turbulence.
\newblock \emph{arXiv preprint arXiv:2501.12740}, 2025.

\bibitem[Lahellec et~al.(2003)Lahellec, Michel, Moulinec, and Suquet]{lahellec2003analysis}
Noël Lahellec, Jean-Claude Michel, Hervé Moulinec, and Pierre Suquet.
\newblock Analysis of inhomogeneous materials at large strains using fast {Fourier} transforms.
\newblock \emph{IUTAM Symposium on Computational Mechanics of Solids Materials}, 108:\penalty0 247--258, 2003.

\end{thebibliography}

\end{document}